\documentclass[journal]{IEEEtran}
\usepackage{amsmath,amsfonts}
\usepackage{algorithmic}
\usepackage{epsfig, algorithm}
\usepackage{array}
\usepackage{amssymb}
\usepackage{acronym} 
\usepackage[caption=false,font=normalsize,labelfont=sf,textfont=sf]{subfig}
\usepackage{textcomp}
\usepackage{stfloats}
\usepackage{tabularx}
\usepackage{xcolor}
\usepackage{url}
\usepackage{verbatim}
\usepackage{multirow}
\usepackage{acronym}
\usepackage{graphicx}	
\usepackage{stfloats}
\usepackage{soul}
\usepackage{cancel}
\def\BibTeX{{\rm B\kern-.05em{\sc i\kern-.025em b}\kern-.08em
T\kern-.1667em\lower.7ex\hbox{E}\kern-.125emX}}
\usepackage{balance}
\usepackage{multirow}
\usepackage{hhline}

\usepackage{easyReview}
\usepackage{cancel}
\begin{document}
\title{Temporal Broadening-Aware Multiplexing for Joint Sensing and Communication in THz Band}
\author{Saira Rafique, Ahmed Naeem, Hüseyin~Arslan~\IEEEmembership{Fellow,~IEEE}
\thanks{Saira Rafique is with the Department of Electrical and Electronics Engineering, Istanbul Medipol University, Istanbul, 34810, Turkey. She is also with the IPR and License Agreements Department, Vestel Electronics, 45030
Manisa, Türkiye (email: saira.rafique@vestel.com.tr).}
\thanks{Ahmed Naeem is with the School of Electrical Engineering and Computer Science National University of Sciences and Technology, Islamabad, Pakistan (email: ahmed.naeem@seecs.edu.pk).}
\thanks{Hüseyin Arslan is with the Department of Electrical and Electronics Engineering, Istanbul Medipol University, Istanbul, 34810, Türkiye (email: huseyinarslan@medipol.edu.tr).}}

\markboth{Journal of \LaTeX\ Class Files,~Vol.~14, No.~8, August~2025}%
{Shell \MakeLowercase{\textit{et al.}}: A Sample Article Using IEEEtran.cls for IEEE Journals}

\maketitle
\begin{abstract} 
High-resolution wireless sensing has become an integral component of futuristic 6G networks alongside high-rate communication. Terahertz (THz) band enables both functions through its extremely large bandwidth, providing sub-centimeter level sensing precision and multi-gigabit data rates. However, THz propagation suffers from severe channel impairments such as molecular absorption (MoA) and the resulting temporal broadening effect (TBE). For sensing, TBE causes temporal spreading of received echoes, leading to degraded range resolution and necessitating long guard intervals between consecutive sensing pulses to avoid overlap. These guards, while necessary for far sensing receiver ($\text{Rx}_{\text{sens}}$), cause latency and inefficient temporal use. To overcome this limitation, this paper proposes a TBE-aware multiplexing framework that exploits the distance-dependent nature of TBE to enable interference-free coexistence of sensing and communication (S\&C) pulses. A guard interval preallocated for the worst-case TBE at far $\text{Rx}_{\text{sens}}$ is opportunistically reused to embed a low-power single-carrier communication pulse for a nearby user experiencing minimal broadening. Limited TBE confines S\&C pulses within their designated slots at short distances, while the broadened and attenuated communication pulse at the distant $\text{Rx}_{\text{sens}}$ becomes negligible, eliminating the need for successive interference cancellation (SIC). Simulation results reveal that compared with power-domain non-orthogonal multiple access (PD-NOMA) and fixed-guard alternative, the proposed scheme achieves superior bit-error rate, sensing accuracy, and latency performance, with up to 66.5\% latency reduction under heavy traffic.
\end{abstract}
\section{Introduction}
Wireless sensing has become an integral component of next-generation wireless communication systems. Beyond data transmission, future 6G networks are expected to perceive and understand their surroundings, enabling advanced capabilities such as high-precision localization, object tracking, gesture and posture recognition, and environment mapping for connected autonomous vehicles (CAVs), smart transportation, and industrial automation \cite{11045549}. Achieving such fine-grained perception requires high-resolution sensing, which directly depends on the signal bandwidth and carrier frequency. Terahertz (THz) band (0.1-10 THz) offers an unprecedented opportunity to realize this goal. Its ultra-wide bandwidth allows sub-centimeter ranging accuracy and millimeter-level spatial resolution, supporting applications that demand simultaneous communication and sensing \cite{9681870}. However, the propagation characteristics of the THz spectrum differ fundamentally from those of millimeter-wave and microwave bands~\cite{11075517}. In particular, molecular absorption (MoA) and the associated temporal broadening effect (TBE) become dominant \cite{han2022terahertz}. MoA leads to frequency-selective attenuation around molecular resonance frequencies (notably water vapor and oxygen), causing strong path-dependent losses while TBE manifests as distance-dependent pulse broadening that degrades signal integrity.
\par For sensing, TBE induces temporal spreading of the received echo even under line-of-sight (LoS) conditions, causing partial overlap between consecutive sensing pulses, leading to range ambiguity and degraded resolution. To preserve sensing accuracy, large guard durations must be inserted between consecutive pulses, particularly for a far sensing receiver ($\text{Rx}_{\text{sens}}$) where TBE is severe. This, however, leads to inefficient temporal utilization. Meanwhile, in many practical scenarios, the same transceiver may serve both a far $\text{Rx}_{\text{sens}}$ (for environmental perception) and a near user equipment ($\text{UE}_{\text{comm}}$) requiring low-latency communication. This raises an important question: can the temporal guard required for sensing be reused to transmit low-power data to a nearby $\text{UE}_{\text{comm}}$ without disturbing sensing performance? This question forms the basis of the broader research direction known as joint sensing and communication (JSAC), which seeks to unify radar-like sensing and data transmission within shared radio resources. JSAC frameworks promise improved spectral efficiency and low-latency operation; however, their performance in THz environments remains constrained by MoA and TBE, which complicate interference management and resource reuse.
\par While several studies have addressed the impact of MoA and TBE on THz communication performance, their focus has remained on improving data reliability rather than enabling simultaneous sensing and communication (S\&C). For instance, the distance-aware multi-carrier (DAMC) scheme adapts modulation and power allocation across frequency sub-windows according to distance-dependent THz channel characteristics, achieving notable data-rate gains under severe attenuation \cite{6884190}. Similarly, the distance-adaptive absorption-peak modulation (DA-APM) method exploits MoA peaks as controllable propagation boundaries, improving transmission secrecy without external jamming \cite{9271892}. In addition, a deep-neural-network-aided receiver has been proposed to reconstruct desired signals under artificial noise, eliminating the need for successive interference cancellation (SIC) and providing robust, low-complexity operation in noisy THz environments \cite{9497766}. Although these techniques effectively enhance communication efficiency and robustness under THz-specific impairments, they do not tackle the core challenge of interference-free coexistence between S\&C. 
\par Existing JSAC frameworks such as time-division JSAC (TD-JSAC) ensure isolation by assigning separate time slots for each function but at the cost of higher latency and inefficient temporal use  \cite{9728752,10273387,10622748}. Power-domain non-orthogonal multiple access (PD-NOMA) between S\&C with SIC, on the other hand, allows concurrent transmission but requires complex signal separation and suffers from residual interference \cite{10697321,10529733}. Additionally, PD-NOMA does not exploit propagation-induced characteristics of the THz channel and therefore represents a channel-agnostic coexistence strategy. To overcome these limitations, this paper introduces a TBE-aware multiplexing framework for THz-band JSAC. The proposed approach leverages the distance-dependent asymmetry of TBE: sensing pulses experience large broadening at far receivers, while near communication signals remain temporally compact. By embedding a low-power single-carrier (SC) communication pulse for the near $\text{UE}_{\text{comm}}$ within the guard interval of a wideband linear frequency modulated (LFM) sensing pulse intended for far $\text{Rx}_{\text{sens}}$, the system enables interference-free coexistence without requiring SIC. The sensing operation retains its wide bandwidth for high-resolution detection through intra-pulse modulation, while the communication link utilizes a SC waveform that achieves moderate-to-high throughput within the THz bandwidth, scalable with higher-order modulation under favorable signal-to-noise-ratio (SNR) conditions.
Furthermore, the framework naturally extends toward hybrid time/frequency-domain multiplexing, where S\&C occupy partially disjoint bandwidths while maintaining temporal isolation \cite{wu2025integrated}. This motivates practical short-range THz deployment scenarios where a single node must support the coexistence of high-rate communication and high-resolution sensing. For instance, a nearby user may require ultra-high data rates, while sensing targets are located farther. Similar configurations arise in industrial and vehicular settings. In such scenarios, distance-dependent temporal broadening creates asymmetric dispersion across receivers, which can be exploited to enable interference-free coexistence without SIC. This improves flexibility and supports dynamic allocation depending on sensing distance and user location.
\begin{figure*}[h]
\centering
\resizebox{1.8\columnwidth}{!}{
\includegraphics{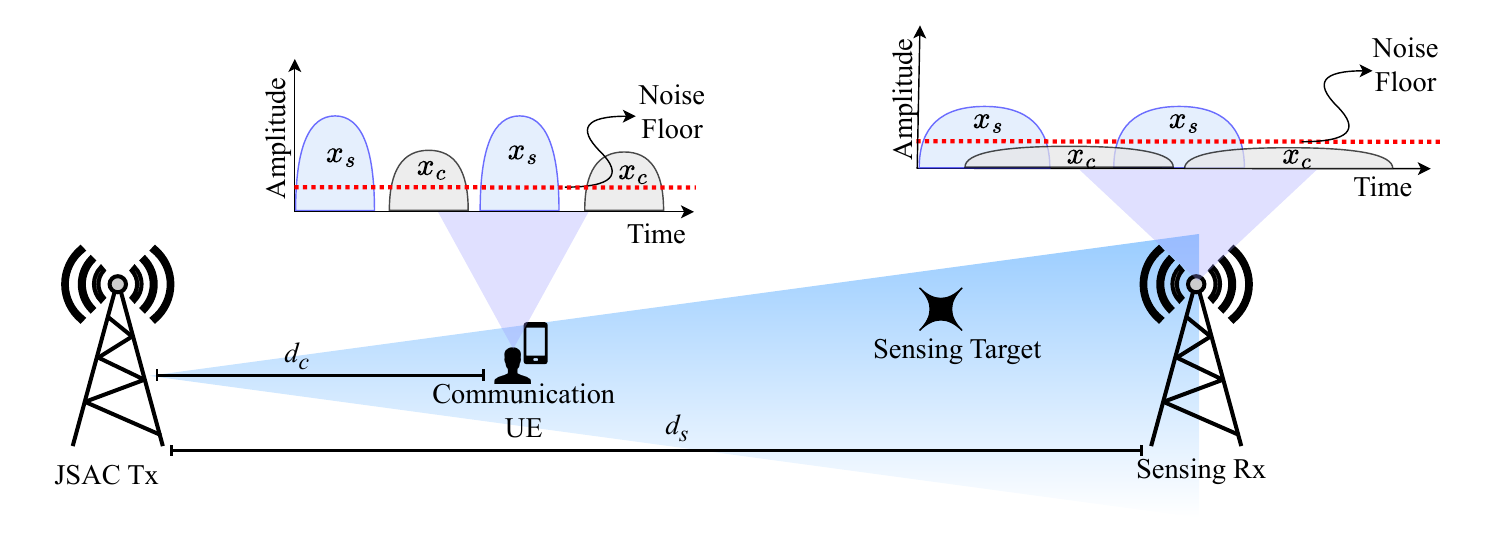}}
\caption{TBE-assisted multiplexing of S\&C signals.}
\label{system_model}
\end{figure*}
The key contributions are summarized as follows.
\begin{itemize}
\item A temporal multiplexing scheme is proposed that opportunistically embeds a low-power communication pulse for a nearby $\text{UE}_{\text{comm}}$ within the guard interval of a high-power sensing pulse directed toward distant $\text{Rx}_{\text{sens}}$. The design exploits the distance-dependent characteristics of TBE, ensuring natural temporal separation of both pulses at their respective receivers.
\item At the near $\text{UE}_{\text{comm}}$, limited TBE ensures that the S\&C pulses remain confined within their allocated intervals. An inner guard is introduced to prevent sensing pulse leakage, and a matched filter is applied to the Gaussian-shaped received signal, enabling low-latency and reliable demodulation without requiring SIC.
\item At far $\text{Rx}_{\text{sens}}$, low-power communication pulse experiences severe TBE and strong attenuation, reducing its power below the noise floor, enabling interference-free sensing without the need for SIC.
\item The communication performance is analyzed in terms of bit error rate (BER) and throughput. Simulation results show that the proposed scheme outperforms its PD-NOMA counterpart, where residual interference after SIC degrades BER. Throughput can be further enhanced by adopting higher-order modulation formats under high SNR conditions.
\item The sensing performance is evaluated using probability of detection ($P_D$) and root-mean-square error (RMSE) of range and velocity, demonstrating superior accuracy compared to PD-NOMA.
\item Additionally, an optimization problem is formulated to determine the interference-free optimal range of distances of the $\text{UE}_{\text{comm}}$ relative to the JSAC transmitter (JSAC Tx), ensuring maximized SINR and reliable coexistence. Frame-level latency is also analyzed using M/D/1 queuing theory and compared with the fixed-guard alternative of the proposed system.
\item Simulation results confirm that the proposed TBE-aware JSAC scheme maintains robust operation across a wide range of distances without requiring power control or complex SIC. Latency analysis further shows up to \textit{66.5\% reduction} compared with the fixed-guard alternative, highlighting the temporal efficiency and scalability of the proposed design.
\end{itemize}
The remainder of the paper is organized as follows. Section \ref{section1} explains the system model of the proposed THz-JSAC system and details the THz channel, and received signal processing. Section \ref{section2} highlights performance evaluation, Section \ref{section3} delineates simulation results and discussions and finally conclusions are drawn in Section \ref{section5}. 
\section{System Model}\label{section1}
\par Consider a JSAC system operating in the THz band, as illustrated in Fig.~\ref{system_model}. The system comprises a JSAC Tx that simultaneously serves a near  $\text{UE}_{\text{comm}}$ and performs bistatic sensing of a far target $L$. The near $\text{UE}_{\text{comm}}$ is located at a distance $d_c$ from JSAC Tx, enabling short-range communication. Meanwhile, the sensing functionality is realized through the target-reflected Tx signal, which is captured at a dedicated sensing receiver ($\text{Rx}_{\text{sens}}$) located at a distance $d_s$ from JSAC Tx, with $d_s > d_c$\footnote{The proposed scheme works most naturally when 
$d_s > d_c$, where sufficient spacing between the two receivers allows the guard induced by TBE at the farther node to be reused without interference. When both receivers are at similar distances or both lie far from the transmitter, the guard may be insufficient. In such cases, TBE–aware pulse-width adaptation can be applied \cite{naeem2025temporal} to confine the near-user signal within its symbol duration, ensuring interference-free S\&C coexistence.}. The spatial separation between $\text{UE}_{\text{comm}}$ and $\text{Rx}_{\text{sens}}$ allows the system to decouple communication and sensing functions through a temporally multiplexed signal~design.
\par To accurately characterize the propagation behavior of THz signals between the JSAC Tx, $\text{UE}_{\text{comm}}$ and $\text{Rx}_{\text{sens}}$, we next describe the underlying THz channel model that incorporates both spreading and MoA losses. Notably, the frequency-selective MoA not only attenuates the signal but also induces TBE, a phenomenon that is leveraged in this work to enable efficient temporal multiplexing of S\&C signals.
\subsection{THz Channel Model}
\par We consider a THz free space line-of-sight channel $H$ composed of spreading loss and MoA loss, consistent with \cite{gao2019distance,9271892}. The complex baseband channel transfer function at frequency $f$ and link distance $d$ is given by
\begin{equation}\label{mainCHANN}
    H(f;d)= H_{\text{abs}}(f;d) H_{\text{spr}}(f;d) e^{-j2\pi f \tau},
\end{equation}
where $\tau=d/c$ and $c$ is the speed of light. In the proposed system model as demonstrated in Fig.~\ref{system_model}, there are two links; the communication link to the $\text{UE}_{\text{comm}}$ at distance $d_c$ and the bi-static sensing link to $\text{Rx}_{\text{sens}}$ at distance $d_s$. Accordingly,
\begin{equation}
    H(f)\in\big\{\,H_{\text{Tx,UE}}(f),\, H_{\text{b}}(f)\,\big\}
    \;=\;\big\{\,H(f;d_c),\, H(f;d_s)\,\big\}.
\end{equation}
\noindent The spreading loss in \eqref{mainCHANN} is given by
\begin{equation}\label{spr}
    H_{\text{spr}}(f;d)=\frac{c}{4\pi f\, d}.
\end{equation}
The MoA loss is modeled by the Beer–Lambert law \cite{5995306,jornet2010channel} as
\begin{equation}\label{expoterm}
    H_{\text{abs}}(f;d)=\exp\Big(-\tfrac{1}{2}k(f) d\Big),
\end{equation}
where $k(f)$ is the medium absorption coefficient. Let $k(f)=\sum_{i,g} k^{i,g}(f)$, where $i$ indexes the isotopologue of gas $g$. For pressure $p$ and temperature $T$,
\begin{equation}\label{chani}
    k^{i,g}(f) \,=\, \frac{p}{p_{0}}\,\frac{T_{\text{STP}}}{T}\; Q^{i,g}\; \sigma^{i,g}(f),
\end{equation}
where $p_0$ and $T_{\text{STP}}$ denote standard pressure and temperature, $Q^{i,g}$ is the molecular density $(\text{molecules}/\text{m}^3)$, and $\sigma^{i,g}(f)$ is the absorption cross section $(\text{m}^2/\text{molecule})$. From the ideal gas law,
\begin{equation}
    Q^{i,g} \,=\, \frac{n}{V}\, q^{i,g} N_A \,=\, \frac{p}{R T}\, q^{i,g} N_A,
\end{equation}
where $q^{i,g}$ is the mixing ratio of isotopologue $i$ of gas $g$, $N_A$ is Avogadro's constant, $n$ is the total number of moles in the mixture, $V$ is the volume, and $R$ is the universal gas constant. The cross section can be rewritten as $\sigma^{i,g}(f)= S^{i,g} G^{i,g}(f)$, where $S^{i,g}$ is the line intensity and $G^{i,g}(f)$ is the spectral line shape. The line center depends on pressure through a linear shift 
$f_c^{i,g} = f_{c0}^{i,g} + \delta^{i,g}\frac{p}{p_0}$,
with $f_{c0}^{i,g}$ the zero-pressure resonance frequency and $\delta^{i,g}$ the pressure shift coefficient~\cite{gordon2022hitran2020}. The MoA occurs over a range of frequencies rather than a single frequency, with the broadening characterized by Lorentz half-width $\alpha_{L}^{i,g}$ given by~\cite{jornet2010channel} 
\begin{equation}
    \alpha_{L}^{i,g}=\Big[(1-q^{g})\alpha_{0}^{\text{air}} + q^{g}\alpha_{0}^{i,g}\Big]
    \Big(\frac{p}{p_0}\Big)\Big(\frac{T_0}{T}\Big)^{\gamma_{\text{temp}}},
\end{equation}
where $\alpha_{0}^{\text{air}}$ and $\alpha_{0}^{i,g}$ are the air-broadened and self-broadened half widths at reference temperature $T_0$, and $\gamma_{\text{temp}}$ is the temperature broadening coefficient. In the THz range the asymmetric Van Vleck–Weisskopf model provides an accurate absorption line shape \cite{jornet2010channel} expressed as
\begin{equation}
\resizebox{0.4\textwidth}{!}{$
    F^{i,g}(f) = \frac{\alpha_{L}^{i,g}}{\pi}\frac{f}{f_c^{i,g}}
    \left[
    \frac{1}{(f-f_c^{i,g})^{2}+(\alpha_{L}^{i,g})^{2}} +
    \frac{1}{(f+f_c^{i,g})^{2}+(\alpha_{L}^{i,g})^{2}}
    \right]$}.
\end{equation}
Continuum absorption is incorporated using a correction at the far ends of the line shape, which yields
\begin{equation}
    G^{i,g}(f) = \frac{f}{f_c^{i,g}}
    \frac{\tanh\!\big(\tfrac{h c f}{2 k_B T}\big)}{\tanh\!\big(\tfrac{h c f_c^{i,g}}{2 k_B T}\big)}
    F^{i,g}(f),
\end{equation}
where $h$ and $k_B$ are Planck and Boltzmann constants. Putting the above together yields
\begin{equation}
    k(f) = \sum_{i,g}\frac{p}{p_0}\frac{T_{\text{STP}}}{T}
    \Big(\frac{p}{R T} q^{i,g} N_A\Big) S^{i,g} G^{i,g}(f),
\end{equation}
and the absorption loss follows from \eqref{expoterm}. The resulting channel is highly frequency selective with pronounced spectral notches and dispersion in bands near molecular resonances \cite{gordon2022hitran2020}. For the communication link and the bi-static sensing link we evaluate \eqref{mainCHANN} at $d=d_c$ and $d=d_s$, respectively as
\begin{equation}
\begin{aligned}
    H_{\text{Tx,UE}}(f)&=H_{\text{abs}}(f;d_c)H_{\text{spr}}(f;d_c)e^{-j2\pi f\tau(d_c)},\\
    H_{\text{b}}(f)&=H_{\text{abs}}(f;d_s)H_{\text{spr}}(f;d_s)e^{-j2\pi f\tau(d_s)}.
\end{aligned}
\end{equation}
The MoA induced dispersion manifests in time as temporal broadening of wideband pulses, which we exploit in the proposed transmit structure to enable interference free temporal multiplexing of S\&C signals within one frame. Building upon the characterized THz channel behavior, particularly the MoA-induced TBE, we now detail the structure of the proposed transmit signal structure that leverages this TBE to enable interference-free temporal multiplexing of S\&C signals within a single frame.
\subsection{Proposed TBE-Aware Signal Structure}
\par
\begin{figure}
\centering
\resizebox{1\columnwidth}{!}{
\includegraphics{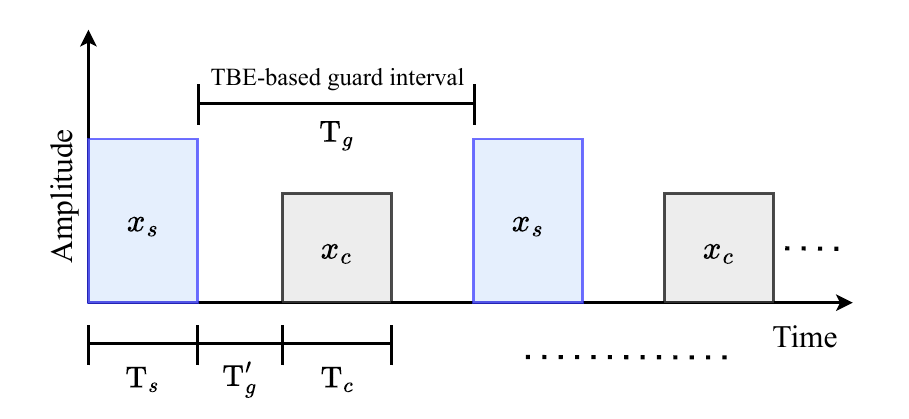}}
\caption{Proposed TBE-aware JSAC signal structure with guard design.}
\label{tbefigi}
\end{figure}

 The JSAC Tx periodically transmits a composite signal consisting of a sensing pulse followed by a communication pulse as shown in Fig.~\ref{tbefigi}. The baseband transmit signal $x(t)$ for one transmission frame is defined as
\begin{equation}
\resizebox{0.48\textwidth}{!}{$
x(t) =
\begin{cases}
\sqrt{P_s} \cdot x_s(t), & 0 \leq t < T_s, \\
\sqrt{P_c} \cdot x_c(t - T_s - T'_g), & T_s + T'_g \leq t < T_s + T'_g + T_c, \\
0, & \text{otherwise},
\end{cases}$}
\end{equation}
where $x_s(t)$ and $x_c(t)$ denote the modulated S\&C pulses, respectively, with associated transmit powers $P_s$ and $P_c$. Typically, $P_s > P_c$, since the sensing operation involves reflections from a distant target, necessitating higher transmit power to receive significant reflections at $\text{Rx}_{\text{sens}}$. Conversely, $\text{UE}_{\text{comm}}$ is located in close proximity to the JSAC Tx, enabling reliable signal demodulation with lower transmit power. 
The sensing pulse $x_s(t)$ is designed as an LFM chirp defined as
\begin{equation} 
x_s(t) = \cos\left(2\pi f_L t + \frac{1}{2} \alpha t^2 \right) \cdot q(t),
\end{equation}
where $f_L$ and $f_H$ denote the start and end frequencies of chirp, respectively, and $\alpha = \frac{B_s}{T_s}$ is chirp rate with $B_s=\left |f_{H}-f_{L} \right|$. The window function $q(t)$ is a rectangular window of duration $T_s$ as
\begin{equation}
q(t) =
\begin{cases}
1, & t \in \left[ -\dfrac{T_s}{2},\ \dfrac{T_s}{2} \right], \\
0, & \text{otherwise}.
\end{cases}
\end{equation}
\par The transmitted baseband communication pulse \( x_c(t) \) is defined as
\begin{equation} 
x_c(t) = a \cdot \text{rect}\left(\frac{t - T_c/2}{T_c}\right),
\end{equation}
where $a \in \{\pm1\}$ is the BPSK symbol and $\text{rect}(\cdot)$ is a unit-amplitude rectangular pulse of duration $T_c$.
The passband signal is given by
\begin{equation} 
x_c^{\text{(PB)}}(t) = \Re \left\{ x_c(t) \cdot e^{j2\pi f_c t} \right\},
\end{equation}
where \( f_c \) is the carrier frequency. This pulse-based design offers low-complexity and robust operation for short-range quasi-static THz links, where TBE is moderate and symbol overlap is negligible \cite{corre:hal-01993187, 9269933}.
where \( f_c \) is the carrier frequency. This pulse-based design offers low-complexity and robust operation for short-range quasi-static THz links, where TBE is moderate and symbol overlap is negligible \cite{corre:hal-01993187, 9269933}. 
Additionally, it is essential to distinguish intrinsic THz propagation effects from design-dependent waveform parameters in the proposed framework. The TBE exploited in this work is an inherent property of the THz channel arising from MoA and depends primarily on propagation distance and environmental conditions. In contrast, parameters such as $B_s$, $T_s$, $T_c$, $P_s$, and $P_c$ are system design choices. These parameters determine how efficiently the system utilizes the channel-induced TBE, but they do not create the temporal separation themselves. Consequently, the guard intervals introduced in the following subsection are derived from the channel-imposed broadening factors rather than arbitrarily selected scheduling parameters.
\subsubsection{Guard Interval Design} 
To mitigate inter-pulse interference caused by TBE in THz channels, two guard intervals are introduced in the JSAC transmission frame. The outer guard interval \( T_g \) separates successive sensing pulses and is designed based on the maximum broadening experienced at \( \text{Rx}_{\text{sens}} \). The inner guard interval \( T'_g \) separates $x_s(t)$ and $x_c(t)$, ensuring interference-free reception at \( \text{UE}_{\text{comm}} \).
\par Since the $x_c(t)$ is an intra-modulated LFM pulse with large bandwidth and $x_c(t)$ is an SC pulse with comparatively lower bandwidth, $x_s(t)$ undergoes more severe broadening. Therefore, \( T'_g \) must account for the broadened tail of the $x_s(t)$ at the near UE to prevent overlap with the $x_c(t)$. The broadened durations of the $x_s(t)$ at \( \text{UE}_{\text{comm}} \) and \( \text{Rx}_{\text{sens}} \) are expressed as \cite{11075517}
\begin{equation} 
\begin{aligned}
\beta_{\text{sens},N} &= 1 + \eta_{br}\cdot d_c, \\
\beta_{\text{sens},F} &= 1 + \eta_{br}\cdot d_s,
\end{aligned}
\end{equation}
where \( \eta_{br}\) is the broadening coefficient for the sensing pulse (accurate value of \( \eta_{br} \)  can be obtained from HITRAN database) \cite{gordon2022hitran2020}. The parameter $\eta_{br}$ is continuously updated based on environmental conditions (humidity, temperature, pressure) and the corresponding receivers' distances. In this work, we emphasize that $\eta_{br}$ is analytically derived using the HITRAN-based MoA model to validate the proposed scheme, ensuring consistency with prior studies \cite{9271892, 9497766, gao2019distance, jornet2010channel}. The corresponding broadened pulse durations are
\begin{equation} 
\begin{aligned}
T_{\text{sens},N}^{\text{rx}} &= \beta_{\text{sens},N} \cdot T_s, \\
T_{\text{sens},F}^{\text{rx}} &= \beta_{\text{sens},F} \cdot T_s.
\end{aligned}
\end{equation}
\par Thus, the required guard intervals are derived as
\begin{equation}\label{beta} 
\begin{aligned}
T'_g &= (\beta_{\text{sens},N} - 1) \cdot T_s, \\
T_g  &= (\beta_{\text{sens},F} - 1) \cdot T_s.
\end{aligned}
\end{equation}
The communication pulse is scheduled to begin at
\begin{equation} 
T_s + T'_g = \beta_{\text{sens},N} \cdot T_s,
\end{equation}
ensuring that the broadened sensing pulse has sufficiently decayed at \(\text{UE}_{\text{comm}}\) before communication begins. This temporal separation minimizes interference from sensing to communication, particularly when \( \text{UE}_{\text{comm}} \) is at moderate distances. However, at short distances, \( T'_g \) becomes small, and the tail of the sensing pulse may not fully decay before the communication pulse arrives, potentially causing residual interference. In such cases, minimal equalization or interference cancellation may be required. Meanwhile, \( T_g \) ensures that consecutive sensing pulses do not overlap at the far sensing receiver, preserving bistatic sensing accuracy and avoiding false returns.
\subsection{Received Signal Processing}
In the considered THz JSAC system, \( x(t) \) contains both  S\&C pulse, temporally multiplexed with a guard interval. The receiver processing differs for the near communication user equipment (UE) and the far sensing receiver (Rx), based on their respective objectives and signal conditions.
\subsubsection{\textbf{Near Communication UE }}
At the near $\text{UE}_{\text{comm}}$, the received signal comprises two temporally separated components: $x_s(t)$ and $x_c(t)$, separated by a TBE-aware $T'_g$. Due to the short $d_c$, the THz channel $h_{\text{Tx,UE}}(t)$, which captures spreading loss and MoA effects, introduces moderate pulse dispersion. Consequently, the originally rectangular communication pulse undergoes TBE as it propagates through the channel.
To characterize the channel-induced dispersion in the time domain, the THz channel impulse response is approximated by a normalized Gaussian kernel given by \cite{naeem2025temporal, lim2022pulse}
\begin{equation} 
h_{\text{Tx,UE}}(t) = \frac{1}{\sqrt{2\pi} \sigma_h} \exp\left(-\frac{t^2}{2\sigma_h^2}\right),
\end{equation}
where $\sigma_h$ is the standard deviation that models dispersion due to MoA and increases with $d_c$. The received baseband signal is given by
\begin{equation} \label{abcd} 
y_c(t) = \sqrt{P_c} \cdot \left[ x_c(t) * h_{\text{Tx,UE}}(t) \right] + n(t),
\end{equation}
where $*$ denotes convolution and $n(t)$ is additive white Gaussian noise with $n(t) \sim \mathcal{CN}(0, \sigma_n^2)$. The convolution $x_c(t) * h_{\text{Tx,UE}}(t)$ smooths the rectangular pulse into a Gaussian-like pulse. Thus, the received communication pulse can be approximated as \cite{naeem2025temporal, lim2022pulse}
\begin{equation} 
p(t) = \frac{1}{\sqrt{2\pi} \sigma} \exp\left(-\frac{(t - T_c/2)^2}{2\sigma^2}\right),
\end{equation}
where the effective standard deviation $\sigma$ ( $\sigma_h \approx \sigma$ is assumed due to the ultra-short duration of the transmitted pulses, such that channel-induced dispersion dominates the received pulse shape) is related to the broadening factor $\beta_{\text{sens},N}$ as 
\begin{equation} 
\sigma = \frac{\beta_{\text{sens},N} T_c}{2\sqrt{2 \ln 2}}.
\end{equation}
Accordingly, the received baseband signal in communication interval $t \in [T_s + T'_g,\ T_s + T'_g + T_c]$ is approximated as
\begin{equation}  \label{eq:y_rx}
y_c(t) = \sqrt{P_c} \cdot a \cdot p(t) + n(t).
\end{equation}
Matched filtering is applied using $p(t)$ as the template to extract the transmitted symbol. The known pulse shape and relatively static channel conditions over the short-duration pulses allow for low-complexity and energy-efficient demodulation. Additionally, $T'_g$ ensures that any residual energy from the preceding sensing pulse remains below the noise floor, maintaining reliable detection performance at $\text{UE}_{\text{comm}}$.
\subsubsection{\textbf{Far Sensing Receiver}} \label{Seci2}
\par The $\text{Rx}_{\text{sens}}$ is positioned to capture bistatic reflections of the transmitted signal from a distant target $L$. The received sensing signal experiences a two-way propagation delay corresponding to the JSAC Tx-$L$-$\text{Rx}_{\text{sens}}$ path. The composite received signal at $\text{Rx}_{\text{sens}}$ is given~by
\begin{equation} 
\begin{aligned}
y_{\text{sens}}(t) &= \underbrace{\alpha_L \cdot h_{\text{b}}(t - \tau_L) * x_s(t)}_{\text{bistatic target return}}
+\underbrace{h_{\text{Tx, Rx}}(t) * x_c(t - T_g)}_{\text{broad. communication pulse}}\\
&+\ \underbrace{\alpha_{\text{UE}} \cdot h_{\text{UE,Rx}}(t - \tau_{\text{UE}}) * x_s(t)}_{\text{UE reflection}}
+ n(t),
\end{aligned}
\end{equation}
\noindent where \( \alpha_L \) and \( \alpha_{\text{UE}} \) denote the reflection coefficients of target and \( \text{UE}_{\text{comm}} \), respectively, \( h_{\text{b}}(t) \) represents the bistatic THz channel from the JSAC Tx to the target \( L \), and subsequently to \( \text{Rx}_{\text{sens}} \), \( h_{\text{Tx,Rx}}(t) \) denotes direct channel from the JSAC Tx to \( \text{Rx}_{\text{sens}} \), \( h_{\text{UE,Rx}}(t) \) is the channel from \( \text{UE}_{\text{comm}} \) to \( \text{Rx}_{\text{sens}} \), and \( \tau_L \) and \( \tau_{\text{UE}} \) are the round-trip delays corresponding to the target \( L \) and \( \text{UE}_{\text{comm}} \), respectively. Specifically, \( \tau_L \) is expressed as
\begin{equation} 
\tau_L = \frac{d_{\text{Tx},L} + d_{\text{Rx}_{\text{sens}},L}}{c},
\end{equation}
\noindent where \( d_{\text{Tx},L} \) and \( d_{\text{Rx}_{\text{sens}},L} \) denote the distances from the JSAC Tx and the \( \text{Rx}_{\text{sens}} \) to $L$. Furthermore, although $x_c(t)$ is also transmitted during the frame, it undergoes severe TBE and high path loss before reaching the $\text{Rx}_{\text{sens}}$. Consequently, its residual energy is diffusely spread and falls below the noise floor, causing negligible interference to the sensing task. This can be quantified by the received communication pulse energy $E_c^{\text{rx}}$, which satisfies
\begin{equation} 
E_c^{\text{rx}} = \int |h_{\text{Tx,Rx}}(t) * x_c(t - T_g)|^2 dt \ll N_0 B,
\end{equation}
where $N_0$ is the noise spectral density and $B$ is the receiver bandwidth. Therefore, $h_{\text{Tx, Rx}}(t) * x_c(t - T_g) \approx 0$. \par Additionally, reflections from the $\text{UE}_{\text{comm}}$ arrive earlier than the target return (\( \tau_{\text{UE}} < \tau_L \)) and are weaker in power due to lower RCS (\( \alpha_{\text{UE}} \ll \alpha_L \)). These reflections are suppressed via delay gating by processing only the interval $t \in [\tau_L - \Delta,\ \tau_L + \Delta]$, where \( \Delta \) corresponds to the broadened pulse width expected at the sensing receiver. After suppressing these interference terms, the simplified received signal at $\text{Rx}_{\text{sens}}$ becomes
\begin{equation} 
y_{\text{sens}}(t) = \alpha_L \cdot h_{\text{b}}(t - \tau_L) * x_s(t) + n(t).
\end{equation}
Matched filtering is performed at $\text{Rx}_{\text{sens}}$ using a time-reversed and conjugated version of $x_s(t)$ yielding
\begin{equation}  \label{target-matched}
\begin{aligned}
z_{\text{sens}}(t) &= y_{\text{sens}}(t) * x_s^*(-t) \\&= \alpha_L \cdot R_{x_s}(t - \tau_L) * h_b(t) + \eta(t), 
\end{aligned}
\end{equation}
where \( R_{x_s}(\tau) \) denotes the autocorrelation of \( x_s(t) \), and \( \eta(t) = n(t) * x_s^*(-t) \) is the filtered noise with variance \( \sigma_{\eta}^2 = N_0 E_s \), where \( E_s \) is the pulse energy. Range estimation is performed by detecting the peak of the matched filter output as
\begin{equation} 
\hat{\tau_L} = \arg \max_t |z_{\text{sens}}(t)|, \quad \hat{d_{L}} = c \cdot \hat{\tau_L}.
\end{equation}
For velocity estimation, \( N \) consecutive sensing pulses are transmitted periodically with pulse repetition interval \(T_g \), corresponding to the outer guard interval between adjacent \( x_s(t) \) pulses as shown in Fig. 2. The Doppler frequency is extracted via the short-time Fourier transform (STFT) \cite{1194413}. Let $z_{\text{sens}}^{(n)}(t)$ be the matched filter output of the $n$-th pulse given by
\begin{equation} 
Z(f, t) = \sum_{n=1}^{N} z_{\text{sens}}^{(n)}(t) e^{-j 2\pi f n T_{g}}.
\end{equation}
The Doppler frequency and the corresponding velocity are
\begin{equation} 
\hat{f}_D = \arg\max_f |Z(f, t)|, \quad \hat{v}_L = \frac{c\hat{f}_{D}}{2 f_c}.
\end{equation}
\par This formulation confirms that reflections from the target dominate the sensing output due to their stronger return and distinct delay. Meanwhile, the communication signal and incidental UE reflections are naturally suppressed due to temporal separation, lower RCS, and path loss, enabling reliable bistatic detection without explicit interference cancellation. The proposed method is explained in Algorithm \ref{algo}.
\section{Performance Evaluation}\label{section2}
This section evaluates the performance of the proposed THz JSAC system by analyzing the effects of distance-dependent channel characteristics, including path loss and TBE. The performance is analyzed from two perspectives: communication accuracy at $\text{UE}_{\text{comm}}$ and target detection capability at $\text{Rx}_{\text{sens}}$.
\subsection{Communication Performance Analysis}
The communication performance is measured in terms of BER and is derived based on the received signal at $\text{UE}_{\text{comm}}$. Matched filtering is applied to the received signal $y_c(t)$ using $p(t)$. The decision statistic is given by
\begin{equation}  \label{ma}
\alpha = \int_0^{T_c} y_c(t) \cdot p(t) \, dt.
\end{equation}
Substituting \eqref{eq:y_rx} into \eqref{ma} gives
\begin{equation} 
\alpha = \sqrt{P_c} \cdot a \int_0^{T_c} p^2(t) \, dt + \int_0^{T_c} n(t) \cdot p(t) \, dt.
\end{equation}
Since $p(t)$ is energy-normalized, i.e., $\int_0^{T_c} p^2(t) \, dt = 1$, the decision statistic simplifies to $\alpha = \sqrt{P_c} \cdot a + \eta,$ where $\eta = \int_0^{T_c} n(t) \cdot p(t)\, dt \sim \mathcal{N}(0, N_0/2)$ is the filtered noise. The final bit decision is made via thresholding
\begin{equation} 
\hat{a} =
\begin{cases}
+1, & \text{if } \alpha > 0, \\
-1, & \text{otherwise}.
\end{cases}
\end{equation}
The bit error probability is given by \cite{10458884}
\begin{equation} 
P_e = \Pr(\alpha < 0 \mid a = +1) = Q\left( \sqrt{ \frac{2P_c}{N_0} } \right),
\end{equation}
where $Q(x) = \int_{x}^{\infty} \frac{1}{\sqrt{2\pi}} e^{-t^2/2} \, dt$. The received energy per bit is $E_b = \int_0^{T_c} |\sqrt{P_c} \cdot a \cdot p(t)|^2 dt = P_c,$ which allows us to express the BER as
\begin{equation} \label{34} 
P_e = Q\left( \sqrt{ \frac{2E_b}{N_0} } \right).
\end{equation}
To incorporate THz channel effects, $E_b$ is adjusted using the frequency response of the channel at \( f_c \), denoted by \( H_{\text{Tx,UE}}(f_c) \). Under the flat-channel assumption (valid for narrowband signaling), the received energy becomes
\begin{equation}  \label{35}
E_b = P_c T_c \cdot |H_{\text{Tx,UE}}(f_c)|^2,
\end{equation}
where
\begin{equation} \label{36}
|H_{\text{Tx,UE}}(f_c)|^2 = \left( \frac{c}{4\pi f_c d_c} \right)^2 \cdot e^{-k(f_c) d_c},
\end{equation}
Substituting \eqref{36} and \eqref{35} into \eqref{34}, the BER expression is
\begin{equation} 
P_e \approx Q\left( \sqrt{ \frac{2 P_c T_c}{N_0} \left( \frac{c}{4\pi f_c d_c} \right)^2 e^{-k(f_c) d_c} } \right).
\end{equation}
\noindent This closed-form expression captures the impact of spreading loss and MoA on the communication reliability at $\text{UE}_{\text{comm}}$. For wideband pulses or frequency-selective channels, numerical integration of $
E_b = \int |H_{\text{Tx,UE}}(f)|^2 |X_c(f)|^2\, df$ is recommended for accurate BER estimation.\\
\begin{algorithm}[t]
\caption{Proposed TBE-Aware S\&C Multiplexing}
\textbf{Input:} $P_s, P_c, T_s, T_c, d_c, d_s, \eta_{br}, f_c, k(f), N_0$ \\
\textbf{Output:} $\hat{a}_k$, $\hat{d}_L$, $\hat{v}_L$
\begin{algorithmic}[1]
\STATE Compute $\beta_{\text{sens},N}$, $\beta_{\text{sens},F}$  \\
\STATE Set guard intervals $T'_g$, $T_g$ \\
\STATE Transmit $x(t)$ \\
\STATE Receive $y_c(t)$ at near $\text{UE}_{\text{comm}}$
\STATE Perform data demodulation via matched filtering and thresholding
\STATE Receive $y_s(t)$ at far $\text{Rx}_{\text{sens}}$
\STATE Estimate $\hat{d}_L$ and  $\hat{v}_L$ through matched filtering and STFT. 
\STATE \textbf{Return:} $\hat{a}_k$, $\hat{d}_L$, $\hat{v}_L$
\end{algorithmic}
\label{algo}
\end{algorithm}
The sensing performance is evaluated using $P_D$ and the RMSE of range and velocity estimates. Specifically, $P_D$ (under a fixed false-alarm probability) quantifies detection reliability, while RMSE captures estimation accuracy based on the delay and Doppler estimates obtained from matched-filter and STFT outputs. Since the proposed framework does not modify the underlying detection or estimation structure, these metrics provide a sufficient characterization of sensing performance.
\subsection{Probability of Detection of the Sensing Target}
\par In the proposed framework, JSAC Tx emits a time-multiplexed signal consisting of $x_s(t)$ followed by $x_c(t)$ within each frame. The $x_s(t)$ is reflected by $L$ and, potentially, by near $\text{UE}_{\text{comm}}$. However, due to the differences in geometry and RCS, the return from the UE is significantly weaker and arrives earlier than the main target echo. Additionally, the LFM-modulated sensing pulse preserves a sharp autocorrelation peak even under TBE, which enables robust range resolution via matched filtering.
\par The matched filter output from \eqref{target-matched} is used to infer the presence of target by examining the magnitude at the expected bistatic delay \( t = \tau_L \). The detection process is modeled using binary hypothesis testing given by \cite{10930959}
\begin{equation} 
\begin{aligned}
\mathcal{H}_0 &: \hat{z}(t) = n(t), \\ 
\mathcal{H}_1 &: \hat{z}(t) = \mathcal{Y}_{\text{target}}(t) + n(t),
\end{aligned}
\end{equation}
where \( \mathcal{Y}_{\text{target}}(t) \) represents the matched filter response corresponding to $L$. A linear envelope detector is applied to the matched filter output, yielding the decision statistic \( \hat{z} = |z_{\text{sens}}(t)| \). Detection is then performed by comparing this magnitude against a predefined threshold \( \mathcal{T} \) as $\hat{z} \underset{\mathcal{H}_0}{\overset{\mathcal{H}_1}{\stackrel{>}{<}}} \mathcal{T}$ \cite{eaves2012principles}. 
\par Under the null hypothesis \( \mathcal{H}_0 \), where only noise is present, \( \hat{z} \) follows a Rayleigh distribution expressed as
\begin{equation} 
p_{\hat{z}}(\hat{z} \mid \mathcal{H}_0) = \frac{2\hat{z}}{\sigma_n^2} \exp\left(-\frac{\hat{z}^2}{\sigma_n^2}\right), \quad \hat{z} \geq 0.
\end{equation}
Accordingly, the probability of a false alarm is given by \cite{10930959} 
\begin{equation} 
P_{\text{FA}} = \int_{\mathcal{T}}^\infty p_{\hat{z}}(\hat{z} \mid \mathcal{H}_0) \, d\hat{z} = \exp\left(-\frac{\mathcal{T}^2}{\sigma_n^2}\right),
\end{equation}
Solving for the detection threshold yields
\begin{equation} 
\mathcal{T} = \sigma_n \sqrt{-\ln P_{\text{FA}}}.
\end{equation}
\par Under hypothesis \( \mathcal{H}_1 \), the received signal contains both signal and noise, and \( z \) follows a Rician distribution as
\begin{equation} \small
p_{\hat{z}}(\hat{z} \mid \mathcal{H}_1) = \frac{2\hat{z}}{\sigma_n^2} \exp\left(-\frac{\hat{z}^2 + m^2}{\sigma_n^2}\right) I_0\left(\frac{2 \hat{z} m}{\sigma_n^2}\right), \quad \hat{z} \geq 0,
\end{equation}
where \( m = |\mathcal{Y}_{\text{target}}(t)| \) represents the magnitude of the deterministic target return, and \( I_0(\cdot) \) is the zeroth-order modified Bessel function of the first kind \cite{baricz2010generalized}. The probability of detection is defined as
\begin{equation} 
P_D = \int_{\mathcal{T}}^\infty p_{\hat{z}}(\hat{z} \mid \mathcal{H}_1) \, d\hat{z}.
\end{equation}
\par For a fixed false alarm rate \( P_{\text{FA}} \), detection probability can be expressed in terms of the first-order Marcum Q-function as 
\begin{equation} 
P_D = Q_1\left(\sqrt{\frac{2 m^2}{\sigma_n^2}}, \sqrt{-2 \ln P_{\text{FA}}}\right),
\end{equation}
where \( Q_1(a, b) \) quantifies the detection performance given the signal-to-noise ratio (SNR) at $\text{Rx}_{\text{sens}}$ and the desired $P_{\text{FA}}$ level.
\subsection{Optimal Interference-Free Distance Range for Multiplexing}
\par This section investigates the conditions under which interference between $x_c(t)$ and $x_s(t)$ can be avoided, and identifies the optimal range of distances for $\text{UE}_{\text{comm}}$ relative to the JSAC Tx. Suppose $\text{UE}_{\text{comm}}$ is located too close to the JSAC Tx (small $d_c$). In that case, it may experience interference from $x_s(t)$ with high $P_s$, despite the presence of $T_g'$. Since $T_g'$ is designed based on TBE, shorter distances result in less dispersion and a reduced $T_g'$, increasing chance of residual interference between high-power $x_s(t)$ and low-power $x_c(t)$.
Conversely, if $\text{UE}_{\text{comm}}$ is located farther from JSAC Tx and nearer to $\text{Rx}_{\text{sens}}$, interference is avoided as the the low-power $x_c(t)$ experiences significant TBE and path loss, keeping it below the noise floor at far $\text{Rx}_{\text{sens}}$, as highlighted in Section \ref{Seci2}. Therefore, the key design constraint becomes ensuring that $x_s(t)$ does not affect communication performance at  $\text{UE}_{\text{comm}}$ due to an insufficient $T_g'$. This motivates optimizing the range of distances for $\text{UE}_{\text{comm}}$ to balance high communication SNR with a sufficient $T_g'$ to eliminate sensing-induced interference.
\subsubsection{Optimization Problem Formulation}
\par To optimize placement of $\text{UE}_{\text{comm}}$, we maximize communication SNR, while ensuring interference-free temporal multiplexing via sufficient $T_g'$. The optimization problem is formulated as
\begin{subequations} \label{opti} 
\begin{align}
\max_{d_c} \quad \text{SNR}_{\text{comm}} =& \frac{P_c c^2 e^{-k(f) d_c}}{(4\pi f d_c)^2 N_0},  \\
\mathrm{subject~to:}&~~
\text{(\textbf{C1})} \quad \text{SNR}_{\text{comm}} \geq \gamma_{\min}, \\
&~~\text{(\textbf{C2})} \quad P_D \geq P_{D,\min}, \\
&~~\text{(\textbf{C3})} \quad d_c < d_s,
\end{align}
\end{subequations}
\subsubsection{Constraints Analysis}
The primary objective is to maximize $\text{SNR}_{\text{comm}}$ while ensuring reliable JSAC performance. The analysis of each constraint is as follows
\par \textbf{Constraint \textbf{C1}} ensures that $\text{SNR}_{\text{comm}}$  is above a threshold \(\gamma_{\min}\) for reliable demodulation. The SNR is defined as
\begin{equation}  \label{idk}
\frac{P_c c^2 e^{-k(f) d_c}}{(4 \pi f d_c)^2 N_0} \geq \gamma_{\min}.
\end{equation}
By defining \( A = \frac{P_c c^2}{(4 \pi f)^2 N_0 \gamma_{\min}} \), the inequality in \eqref{idk} becomes
\begin{equation} 
\frac{e^{-k(f) d_c}}{d_c^2} \geq \frac{1}{A}.
\end{equation}
Multiplying both sides by \( d_c^2 e^{k(f) d_c} \), we get
\begin{equation} 
d_c^2 e^{k(f) d_c} \leq A.
\end{equation}
This defines the upper bound \( d_c \leq d_{c,\max} \), where \( d_{c,\max} \) is the solution of the above transcendental inequality. For small \( k(f) d_c \), using \( e^{k(f) d_c} \approx 1 \), we get the approximation as
\begin{equation} 
d_{c,\max} \approx \sqrt{\frac{P_c c^2}{(4 \pi f)^2 N_0 \gamma_{\min}}}.
\end{equation}
\par \textbf{Constraint \textbf{C2}} ensures that \( P_D \) for bistatic sensing remains above a threshold \( P_{D,\min} \). Since the power of $x_s(t)$ satisfies \( P_s \gg P_c \), and the bistatic sensing performance depends mainly on the strength of the reflected sensing pulse and not on the $\text{UE}_{\text{comm}}$ placement \( d_c \) (provided that $P_c$ is fixed), this condition is satisfied as long as
\begin{equation} 
P_D = Q_1\left(\sqrt{\frac{2 m^2}{\sigma_n^2}}, \sqrt{-2 \ln P_{\text{FA}}}\right) \geq P_{D,\min},
\end{equation}
with sufficient \( P_s \) and reasonable bistatic geometry. Thus, \( P_D \) is independent of \( d_c \), and constraint \textbf{C3} does not affect the optimization of $\text{UE}_{\text{comm}}$ placement. However, for reliable communication when power adaptation is applied based on $d_c$, it will deteriorate $P_D$, as shown in Section IV-D.
\par \textbf{Constraint \textbf{C3}} maintains a geometric order such that $\text{UE}_{\text{comm}}$ is located closer to JSAC Tx than $\text{Rx}_{\text{sens}}$, i.e., $d_c < d_s.$ This reflects inherent system design: $\text{UE}_{\text{comm}}$ is near the JSAC Tx, while bistatic sensing is performed over a longer range.
\subsubsection{Optimization of $d_c$}
The objective function from (49a) aims to maximize $\text{SNR}_{\text{comm}}$ 
\begin{equation} \label{ahmi} 
\text{SNR}_{\text{comm}} = \frac{P_c c^2 e^{-k(f) d_c}}{(4\pi f d_c)^2 N_0}.
\end{equation}
To analyze this function, we compute the derivative as
\begin{equation}  \label{behave}
\frac{\text{d}}{\text{d}d_c} \left( \frac{e^{-k(f) d_c}}{d_c^2} \right)
= e^{-k(f) d_c} \left( -\frac{k(f)}{d_c^2} - \frac{2}{d_c^3} \right).
\end{equation}
Since \( k(f) > 0 \) and \( d_c > 0 \), the derivative is negative, indicating that \( \text{SNR}_{\text{comm}} \) monotonically decreases with increasing \( d_c \). Thus, the best SNR is achieved at the smallest feasible distance, \( d_c = d_{c,\min} \). However, reducing \( d_c \) reduces $T_g'$. A shorter \( T'_g \) may be insufficient to isolate the sensing pulse from the communication signal at the UE, introducing interference. Therefore, a trade-off must be maintained between maximizing \( \text{SNR}_{\text{comm}} \) and allocating a sufficient guard interval.
\par To achieve this, we define a utility function that jointly considers \( \text{SNR}_{\text{comm}} \) and guard interval overhead
\begin{align} 
U(d_c) &= w_1 \cdot \text{SNR}_{\text{comm}} - w_2 \cdot T'_g \notag \\
&= w_1 \frac{P_c c^2 e^{-k(f) d_c}}{(4\pi f d_c)^2 N_0} - w_2 \eta_{br} d_c T_s,
\end{align}
where \( w_1, w_2 > 0 \) are weights that prioritize SNR maximization and guard interval minimization, respectively. To explicitly capture the three competing physical phenomena 
governing the placement of $\text{UE}_\text{comm}$, the utility 
function is restructured as
\begin{align} 
U(d_c) = \frac{w_1 P_c c^2}{N_0} \cdot 
          \underbrace{\frac{1}{(4\pi f d_c)^2}}_{\text{Path Loss}} \cdot 
          \underbrace{e^{-k(f)d_c}}_{\text{MoA}} 
        - w_2 \underbrace{\eta_{br} \, d_c \, T_s}_{\text{Guard Overhead}},
\label{newafter}
\end{align}
\noindent where path loss and MoA jointly degrade received signal 
quality as $d_c$ increases, decaying as $d_c^{-2}$ and 
$e^{-k(f)d_c}$, respectively, while guard overhead grows linearly 
through $\eta_{br}$. These opposing behaviors create a fundamental 
tension: smaller $d_c$ improves SNR but reduces TBE-induced temporal 
isolation, whereas larger $d_c$ improves isolation but accelerates 
signal attenuation. In practical deployment, from \eqref{newafter} the $w_1$ is selected according to the minimum required communication SNR $\gamma_{\min}$ (QoS-driven), while $w_2$ reflects the maximum tolerable guard duration imposed by sensing resolution. Taking the derivative of \( U(d_c) \) and setting it to zero yields
\begin{equation} \label{zero} \small
w_1 \frac{P_c c^2}{(4\pi f)^2 N_0} \left( -\frac{k(f) e^{-k(f) d_c}}{d_c^2} - \frac{2 e^{-k(f) d_c}}{d_c^3} \right) - w_2 \eta_{br} T_s = 0.
\end{equation}
Rewriting and simplifying, we get
\begin{equation} \label{vallah} 
\frac{e^{-k(f) d_c}}{d_c^2} \left( k(f) + \frac{2}{d_c} \right) = \frac{w_2 \eta_{br} T_s (4\pi f)^2 N_0}{w_1 P_c c^2}.
\end{equation}
By defining \( C = \frac{w_2 \eta_{br} T_s (4\pi f)^2 N_0}{w_1 P_c c^2} \), the equation becomes
\begin{equation} 
e^{-k(f) d_c} \left( \frac{k(f)}{d_c^2} + \frac{2}{d_c^3} \right) = C,
\end{equation}
\begin{align}
\underbrace{\frac{2\,e^{-k(f)d_c}}{d_c^3}}_{\text{Path Loss rate}} 
+ \underbrace{\frac{k(f)\,e^{-k(f)d_c}}{d_c^2}}_{\text{MoA rate}} 
= \underbrace{C}_{\text{Guard rate}}.
\label{eq:optimality_condition}
\end{align}
\par This transcendental equation can be solved numerically within a feasible range \( d_c \in [d_{c,\min}, d_{c,\max}] \). For small \( k(f) d_c \), we can approximate \( e^{-k(f) d_c} \approx 1 \) and neglect \( \frac{k(f)}{d_c^2} \), yielding
\begin{align} 
\frac{2}{d_c^3} \approx C \quad \Rightarrow \quad d_c \approx \left( \frac{2}{C} \right)^{1/3}.
\end{align}
Substituting back, we get an approximate closed-form solution
\begin{equation} 
d_c \approx \left( \frac{2 w_1 P_c c^2}{w_2 \eta_{br} T_s (4\pi f)^2 N_0} \right)^{1/3}.
\end{equation}
This provides the optimal trade-off point that balances received \( \text{SNR}_{\text{comm}} \) and temporal guard overhead.
\subsubsection{Optimal Distance Range}
The feasible distance range for interference-free communication is defined by
\begin{equation} 
d_{c,\min} = \sqrt{\frac{P_c c^2}{(4\pi f)^2 N_0 B}}, \quad
d_{c,\max} = \sqrt{\frac{P_c c^2}{(4\pi f)^2 N_0 \gamma_{\min}}},
\end{equation}
where $B$ refers to the effective noise bandwidth of the $\text{UE}_\text{comm}$, with additional geometric condition $d_c < d_s.$ The optimal \( d_c \) lies within this range, ensuring maximum achievable SNR at $\text{UE}_{\text{comm}}$, a sufficiently long \( T'_g \) to avoid overlap with the broadened $x_s(t)$, and system feasibility per bistatic geometry. It is worth mentioning here that RMSE is not explicitly included in the optimization formulation, as under the designed interference-free regime ($E_c^{\mathrm{rx}} \ll N_0 B$), sensing accuracy is primarily governed by the sensing waveform and channel conditions rather than the communication link parameters. Therefore, RMSE is used as a post-optimization validation metric.
\subsection{SINR Analysis for Non-Optimal Distance Range}
\par When the $\text{UE}_{\text{comm}}$ is positioned outside the optimal distance range \( d_{c,\min} \leq d_c \leq d_{c,\max} \), the system performance may degrade. In this section, we analyze SINR at the $\text{UE}_{\text{comm}}$.
\par \textbf{Near communication UE} (\( d_c < d_{c,\min} \) or \( d_c > d_{c,\max} \)):
If \( d_c < d_{c,\min} \), the guard interval \( T'_g\) becomes too short due to minimal TBE. This allows the high-power \( x_s(t) \) to overlap with power \( x_c(t) \), introducing interference. On the other hand, if \( d_c > d_{c,\max} \), although \( T'_g \) may be sufficient, the increased path loss significantly reduces the received SINR at the $\text{UE}_{\text{comm}}$. The received signal in \eqref{abcd} in the presence of overlap is rewritten as
\begin{equation} \small
y_c(t) = h_{\text{TX,UE}}(t) * \left[ \sqrt{P_s} \cdot x_s(t) + \sqrt{P_c} \cdot x_c(t - T_s - T'_g) \right] + n(t),
\end{equation}
where the interference from $x_s(t)$ during the overlap is
\begin{equation} 
I_s(t) = \sqrt{P_s} \cdot h_{\text{TX,UE}}(t) * x_s(t).
\end{equation}
The interference power is defined as
\begin{equation} 
P_I = 
\begin{cases}
P_s \cdot |H_{\text{TX,UE}}(f, d_c)|^2 \cdot \left(1 - \dfrac{T'_g}{T_s}\right), & \text{if } T'_g < T_s \\
0, & \text{otherwise}
\end{cases}.
\end{equation}
The resulting SINR at the UE is
\begin{equation}
\text{SINR} = \frac{P_c |H_{\text{TX,UE}}(f, d_c)|^2}{P_I + N_0 B}.
\end{equation}
For \( d_c < d_{c,\min} \), the short \( T'_g \) causes significant temporal overlap, increasing \( P_I \) and degrading SINR. For \( d_c > d_{c,\max} \), \( |H_{\text{TX,UE}}(f, d_c)|^2 \) drops, decreasing resultant SINR even if no overlap occurs. Both scenarios reduce communication reliability. Consequently, the SINR analysis for non-optimal \( d_c \) primarily highlights the impact of improper guard interval scaling on communication performance.
\par \textbf{Far sensing receiver:}  
As discussed earlier, \( x_c(t) \), due to its low power and heavy attenuation from absorption and spreading, does not exceed the noise floor at $\text{Rx}_{\text{sens}}$ even if $\text{UE}_{\text{comm}}$ is positioned close to it. Thus, there is no practical interference at $\text{Rx}_{\text{sens}}$, and its sensing performance remains unaffected by changes in \( d_c \). 
\subsection{Latency Analysis}
\par This section presents the latency analysis of the proposed TBE-aware JSAC system and its fixed-guard counterpart under the same waveform configuration. The latency is defined as the expected time elapsed from the arrival of a communication packet at the transmitter until its transmission begins. It consists of two components: (i) queuing delay due to packet arrival dynamics, and (ii) scheduling delay defined by the frame structure. For both systems, we model communication packet arrivals as a Poisson process with rate $\lambda$ and assume deterministic frame-based service. Therefore, the average latency follows the well-established M/D/1 queuing model \cite{1096634, daley1976queueing}.
In the fixed guard configuration, the communication signal is scheduled after a fixed guard interval $T_g$ that accounts for the worst-case TBE experienced at far $\text{Rx}_{\text{sens}}$. Accordingly, the total frame duration is given by
\begin{equation} 
T_{\text{frame}}^{\text{TD}} = T_s + T_g + T_c = T_s + \eta_{br} d_s T_s + T_c.
\end{equation}
\par The corresponding scheduling delay, defined as the time from the beginning of the frame until the communication pulse, is given by
\begin{equation} 
\Delta_{\text{latency}}^{\text{TD}} = T_s + T_g = \left( 1 + \eta_{br} d_s \right) T_s.
\end{equation}
Under M/D/1 assumptions, average queuing delay is \cite{485740}
\begin{equation} 
W_q^{\text{TD}} = \frac{\rho_{\text{TD}} \left( T_{\text{frame}}^{\text{TD}} \right)^2}{2 \left( 1 - \rho_{\text{TD}} \right)},
\end{equation}
where $\rho_{\text{TD}} = \lambda T_{\text{frame}}^{\text{TD}}$ is the utilization factor. The total average latency is then expressed as
\begin{equation} \small
\mathbb{E}[L_{\text{TD}}] = W_q^{\text{TD}} + \Delta_{\text{latency}}^{\text{TD}} = \frac{\lambda \left( T_{\text{frame}}^{\text{TD}} \right)^2}{2 \left( 1 - \rho_{\text{TD}} \right)} + \left( 1 + \eta_{br} d_s \right) T_s.
\end{equation}
In contrast, the proposed TBE-aware JSAC system dynamically adapts the guard interval $T'_g$ based on the actual TBE experienced by the near $\text{UE}_{\text{comm}}$ located at distance $d_c$. This reduces unnecessary guarding and improves temporal efficiency. The resulting reduced frame duration is given by
\begin{equation} 
T_{\text{frame}}^{\text{TBE}} = T_s + T_g' + T_c = T_s + \eta_{br} d_c T_s + T_c,
\end{equation}
and a corresponding scheduling delay is
\begin{equation} 
\Delta_{\text{latency}}^{\text{TBE}} = T_s + T_g' = \left( 1 + \eta_{br} d_c \right) T_s.
\end{equation}
The average latency in the TBE-aware system is derived as
\begin{equation} 
\mathbb{E}[L_{\text{TBE}}] = \frac{\lambda \left( T_{\text{frame}}^{\text{TBE}} \right)^2}{2 \left( 1 - \rho_{\text{TBE}} \right)} + \left( 1 + \eta_{br} d_c \right) T_s,
\end{equation}
where $\rho_{\text{TBE}} = \lambda T_{\text{frame}}^{\text{TBE}}$. The average latency reduction achieved by the proposed scheme is obtained by subtracting the two expressions
\begin{equation}
\begin{aligned}
\Delta L &= \mathbb{E}[L_{\text{TD}}] - \mathbb{E}[L_{\text{TBE}}]\\& = \frac{\lambda}{2} \left[ \frac{ \left( T_{\text{frame}}^{\text{TD}} \right)^2 }{1 - \rho_{\text{TD}}} - \frac{ \left( T_{\text{frame}}^{\text{TBE}} \right)^2 }{1 - \rho_{\text{TBE}}} \right] + \eta_{br} T_s (d_s - d_c).
\end{aligned} 
\end{equation}
This expression captures the total latency gain arising from both reduced queuing delay and adaptive scheduling. The relative reduction in scheduling delay, also interpretable as a measure of temporal efficiency, is given by
\begin{equation}
\zeta = \frac{\Delta_{\text{latency}}^{\text{TD}} - \Delta_{\text{latency}}^{\text{TBE}}}{\Delta_{\text{latency}}^{\text{TD}}} = 1 - \frac{1 + \eta_{br} d_c}{1 + \eta_{br} d_s}.
\end{equation}
For typical scenarios where $d_c \ll d_s$, this simplifies to~\mbox{$\zeta \approx 1 - \frac{d_c}{d_s}$}.

\section{Simulation Results and Discussions}\label{section3}
\par This section presents detailed simulations to evaluate the performance of the proposed TBE-aware JSAC system across key metrics, including SINR, BER, throughput, $P_D$, RMSE of range and velocity, and communication latency. To facilitate interpretation of the results, the proposed framework is compared with two coexistence strategies. PD-NOMA with SIC represents channel-agnostic multiplexing of S\&C and serves as the primary baseline for evaluating communication reliability and sensing accuracy. In contrast, the fixed-guard TD-JSAC configuration represents conventional orthogonal temporal separation using conservative guard allocation and is used to assess latency performance. The simulation parameters are shown in Table \ref{tablei}.
\begin{table}
\centering
\caption{Simulation parameters}
\renewcommand{\arraystretch}{1.3}
\setlength{\tabcolsep}{3pt}
\begin{tabular}{|p{92pt}|p{85pt}|}
\hline
\textbf{Parameter} & \textbf{Value}  \\
 \hline
 Carrier frequency   & $f_c= 0.1$~THz   \\
 \hline
 Distance to $\text{UE}_{\text{comm}}$ & $d_c=3$ m   \\
 \hline
 Distance to $\text{Rx}_{\text{sens}}$ & $d_s=10$ m   \\
 \hline
 Target range & $d_L=8$ m\\
 \hline
Target velocity & $v_L=0.6$ m/s\\
\hline
Chirp Bandwidth & $B_s=2$ GHz\\
\hline
Broadening coefficient & $\eta_{br}=0.004 \text{m}^{-1}$ \\
 \hline
Pulse Duration  &  $T_s=T_c=500$ ns\\
\hline
Guard duration & $T'_g=6$ ns,  $T_g=20$ ns  \\
\hline
Frame duration & $T_{\text{frame}}^{\text{TBE}}=$ 1026 ns \\
\hline
Sensing pulse power & $P_s=25$ dbm \\
\hline
Communication pulse power & $P_c=10$ dbm \\
\hline
\end{tabular}
\label{tablei}
\end{table}
\subsection{SINR Analysis across $\text{UE}_{\text{comm}}$ Distances}
Figure 3(a) illustrates the variation of SINR with respect to $d_c$ under two scenarios: with and without a distance-adaptive guard interval $T'_g$. At short distances ($d_c < 2\,\text{m}$), SINR is low in both cases but slightly higher with guard. In the absence of guard, strong sensing pulses and weak communication signals temporally overlap due to negligible broadening, resulting in severe interference. With guard, $T'_g$ is small due to the proximity, and although partial temporal isolation is achieved, the high-power $x_s(t)$ still leaks into $x_c(t)$, causing residual interference. As $d_c$ increases to the optimal range of distances ($2\,\text{m} \leq d_c \leq 3.5\,\text{m}$), SINR improves significantly, reaching a peak at $d_c = 3\,\text{m}$. This improvement arises from increased temporal broadening of the sensing pulse and sufficient $T'_g$, which together suppress interference. Furthermore, $x_c(t)$ still retains sufficient strength before path loss becomes dominant. Beyond $d_c > 3.5\,\text{m}$, SINR declines sharply in both cases. In the guard-enabled case, the drop is steeper due to dominance of path loss and MoA, which severely attenuate the already low-powered communication signal despite full interference suppression. Without guard, SINR also degrades due to increased path loss; however, the SINR remains consistently lower across all distances, being doubly affected by both residual interference at short-to-mid ranges and attenuation at longer distances. These findings confirm that distance-adaptive guard mechanism enhances SINR performance most effectively in mid-range scenarios. However, techniques such as power control or range-aware coding may be required to sustain performance at extended distances.
\begin{figure*}[ht]
\centering\subfloat[]{ \includegraphics[width=60mm,height=55mm]{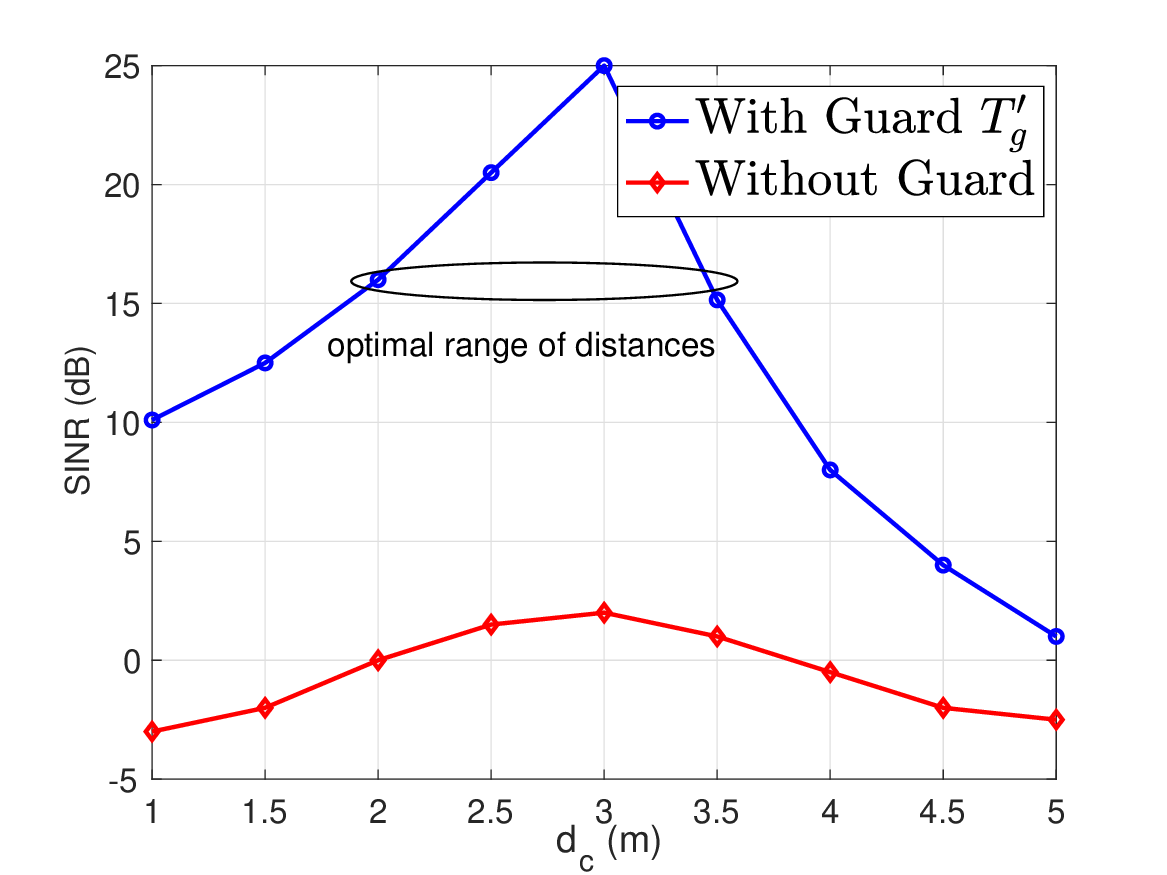}}%
\subfloat[]{ \includegraphics[width=60mm,height=55mm]{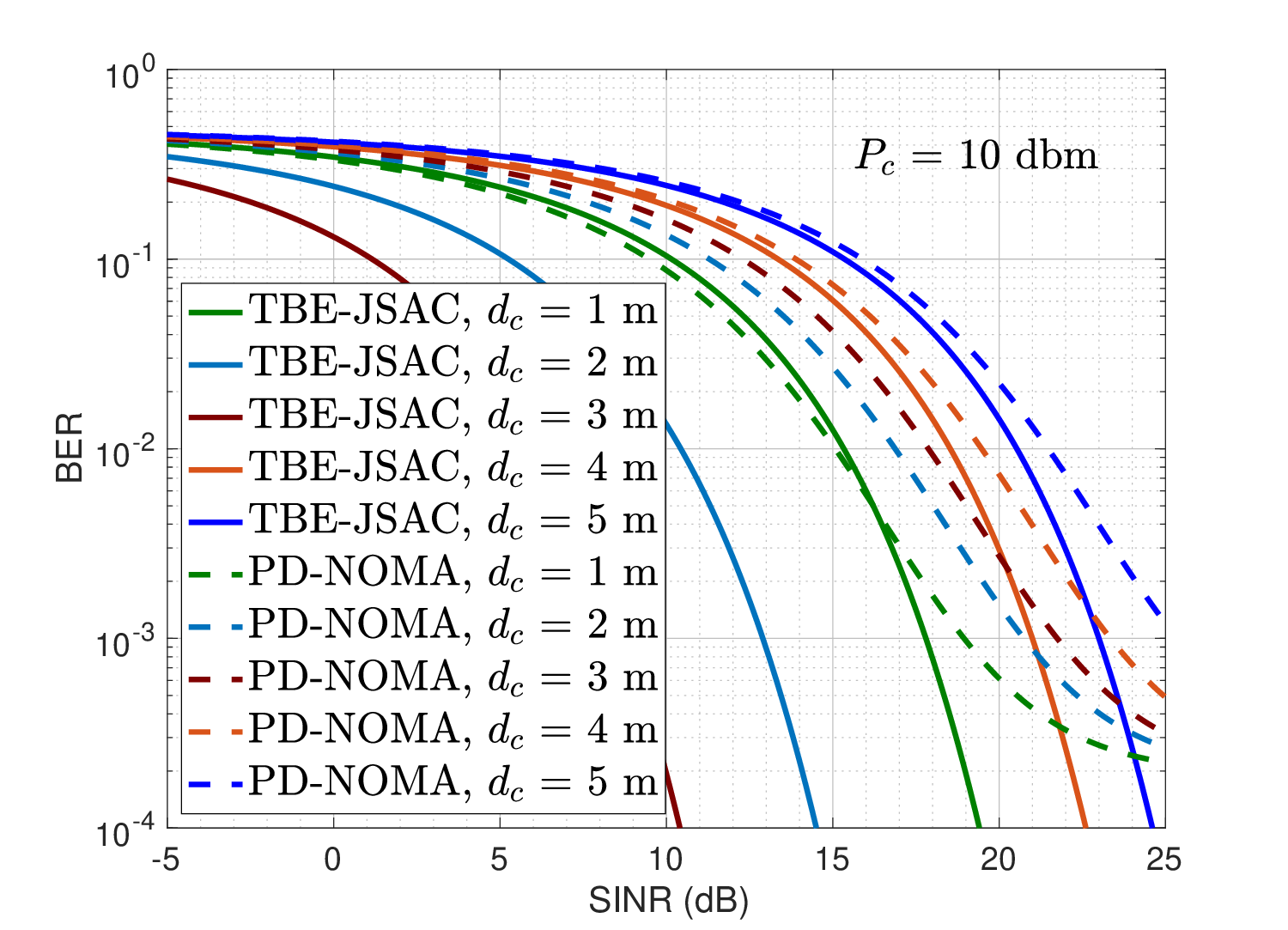}}%
\subfloat[]{ \includegraphics[width=60mm,height=55mm]{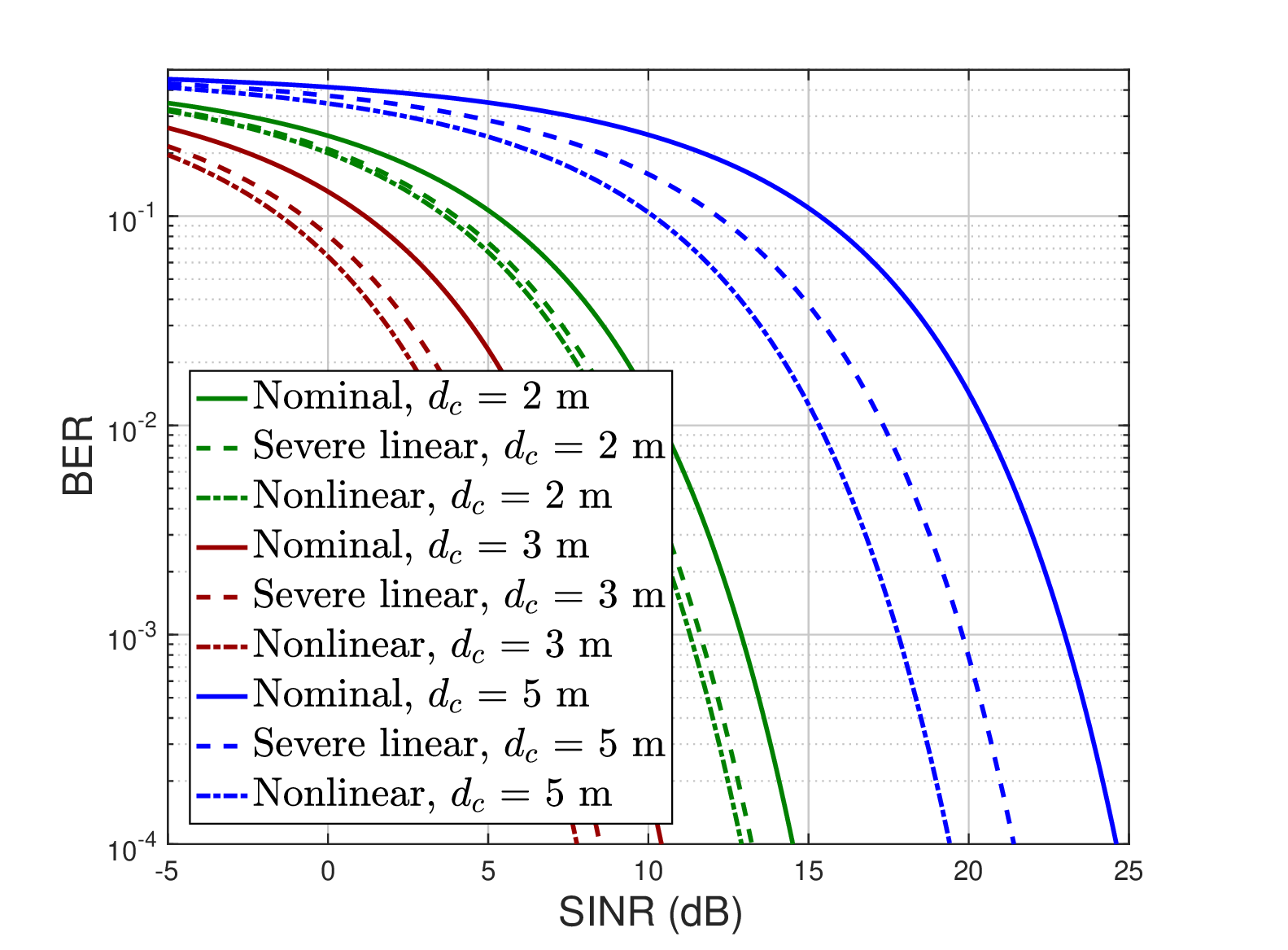}}
\caption{(a) SINR vs $d_c$ with and without guard $T'_g$, (b) BER vs SINR for $\text{UE}_{\text{comm}}$ under TBE-JSAC and PD-NOMA at various $d_c$, and (c) BER versus SINR for $\mathrm{UE}_{\mathrm{comm}}$ at $d_c = 2$, $3$, and $5$~m under nominal, severe-linear, and nonlinear TBE models.}
\label{simres}
\end{figure*}
\begin{figure*}[ht]
\centering
\subfloat[]{\includegraphics[width=60mm,height=55mm]{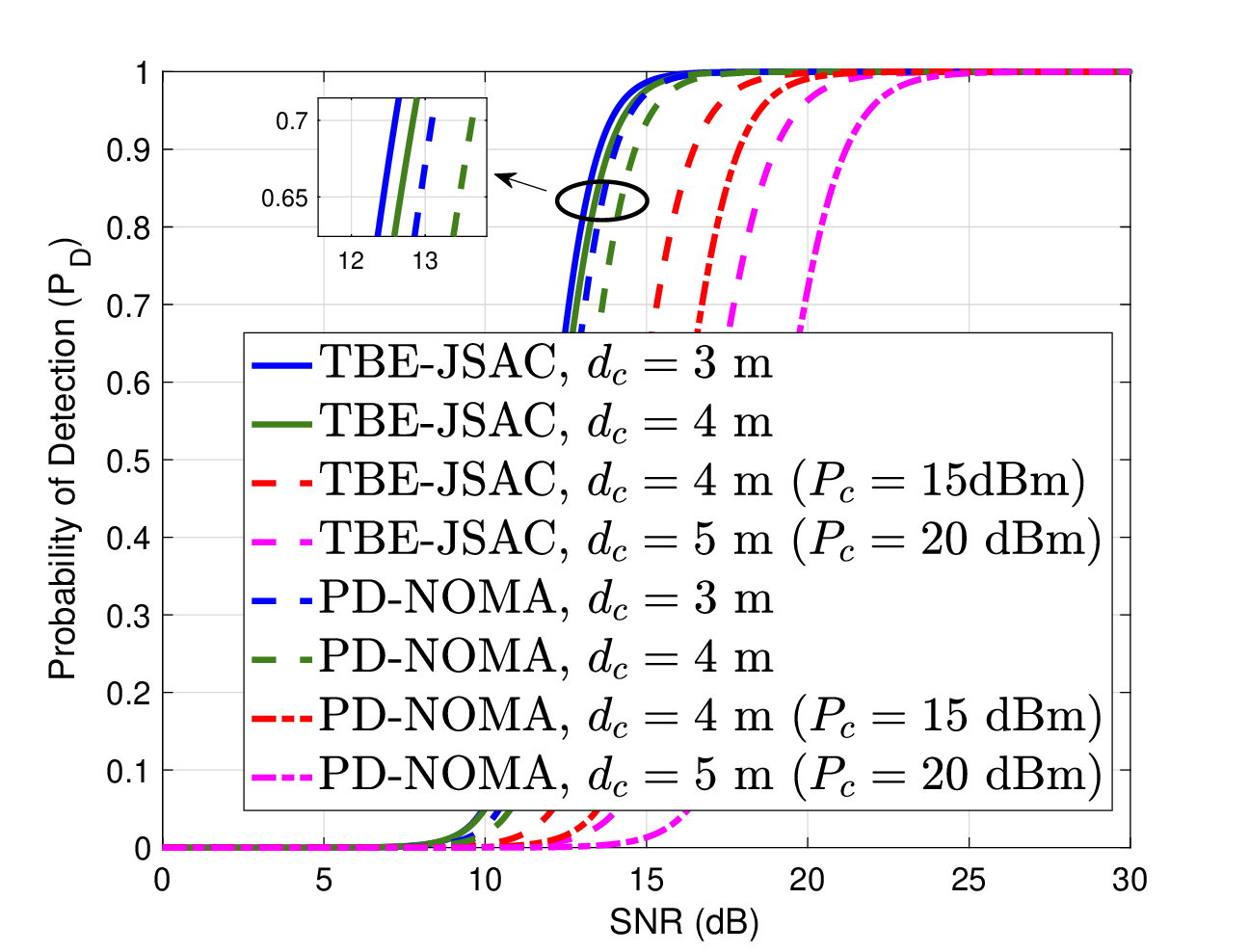}}
\subfloat[]{ \includegraphics[width=60mm,height=55mm]{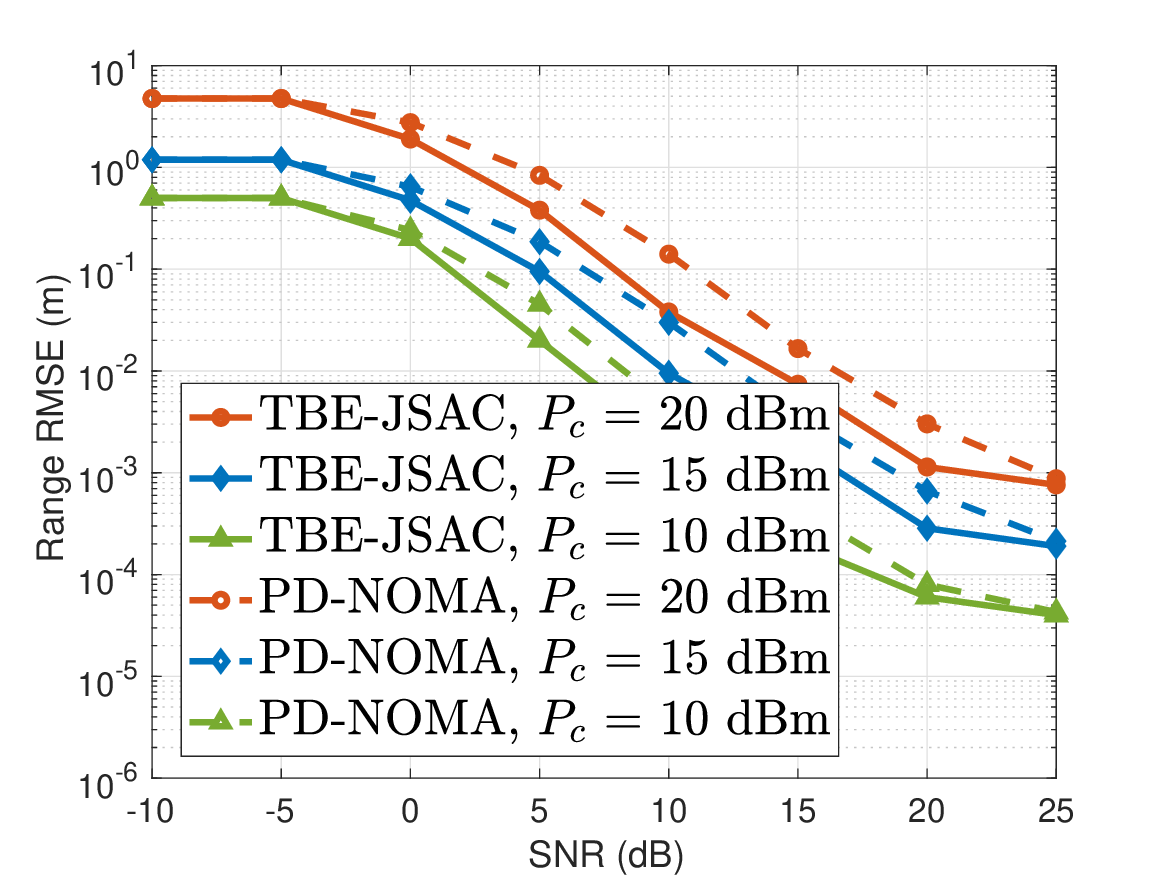}}
\subfloat[]{ \includegraphics[width=60mm,height=55mm]{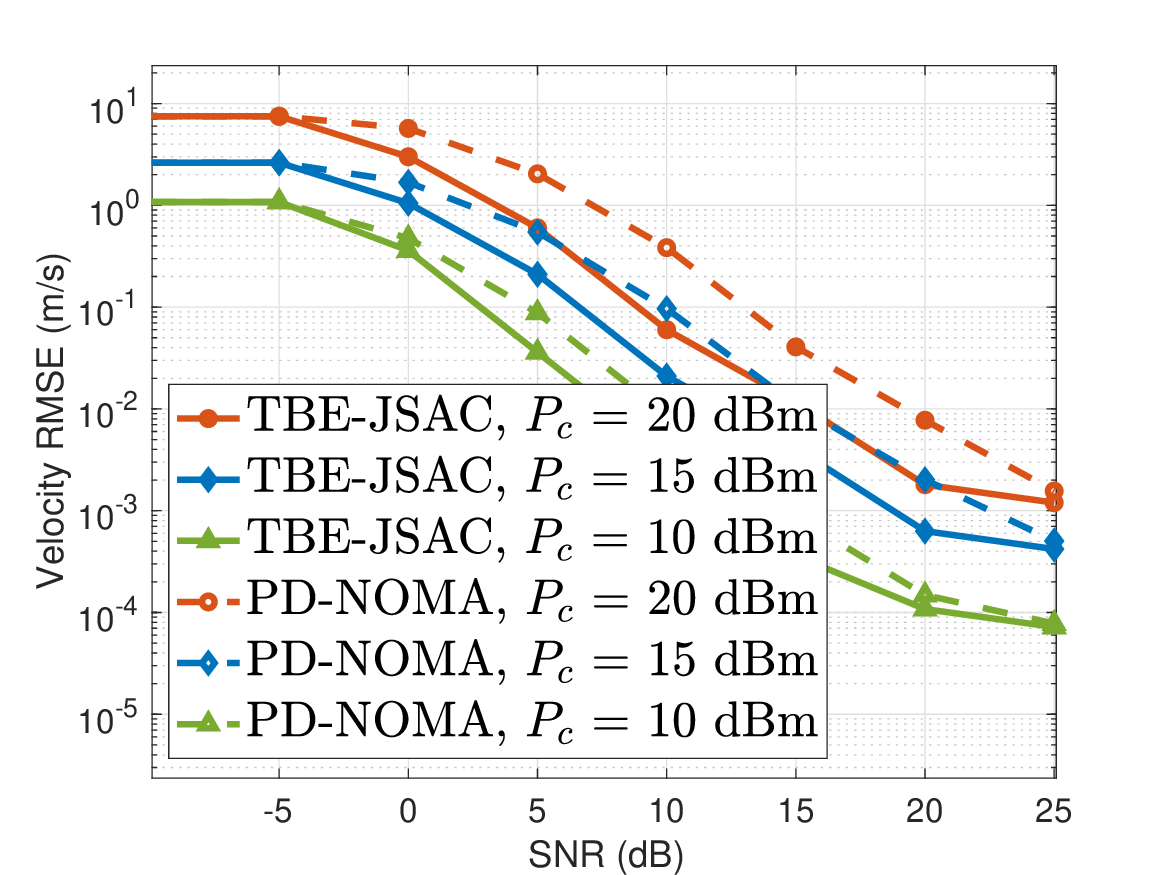}}
\caption{(a) SNR vs $P_D$ for $\text{Rx}_{\text{sens}}$, (b) SNR vs range RMSE at various $P_c$, and (c) SNR vs velocity RMSE at various $P_c$.}
\label{sens_accuracy}
\end{figure*}
\subsection{Bit Error Rate Analysis}
Figure~3(b) illustrates the BER performance across $d_c \in \{1,2,3,4,5\}$m for both the proposed TBE-aware JSAC (solid) and PD-NOMA with SIC (dashed) schemes. For a fair comparison, the same waveform configuration is implemented in both schemes with PD-NOMA employing SIC to separate overlaid signals.
The two approaches exhibit fundamentally different interference behaviors and, consequently, distinct SINR--BER trends. In the TBE-aware JSAC, $T_g'$ increases with distance and, together with the MoA-induced TBE, effectively suppresses the sensing-to-communication leakage power at the $\text{UE}_{\text{comm}}$. As a result, the solid curves improve monotonically with SINR, showing an optimal mid-range ($d_c = 2$--$3$~m) where interference between $x_s(t)$ and $x_c(t)$ is minimized without excessive path loss. At very short distances ($d_c = 1$~m), the guard is insufficient to isolate the high-power sensing pulse, leading to an elevated interference floor and degraded BER, whereas at long distances ($d_c = 4$--$5$~m$)$, the BER degradation stems primarily from link-budget limitations due to spreading loss and MoA. In contrast, the PD-NOMA + SIC curves display a flatter slope and saturation-like behavior at higher SINR values, reflecting the impact of residual interference after imperfect SIC. This residual term introduces a cancellation-limited regime in which further SINR improvements yield only marginal BER gains, particularly for near and mid-range distances. Consequently, PD-NOMA is comparable to the proposed scheme at $d_c = 1$~m, clearly worse at $d_c = 2$--$3$~m, and nearly overlaps at $d_c = 4$--$5$~m where both systems become noise- and path-loss-limited. These observations confirm that the proposed TBE-aware JSAC achieves interference-free operation through distance-dependent temporal isolation, eliminating the need for SIC. 
\subsection{Robustness to Strong TBE}
To further assess the robustness of the proposed framework beyond the nominal quasi-linear TBE regime used in this work. Fig.~3(c) evaluates the BER performance of the TBE-aware JSAC scheme for $d_c = 2$, $3$, and $5$~m, under three TBE profiles. To emulate harsher propagation regimes, two additional cases are considered: (i) a severe linear broadening scenario with an increased $\eta_{br} = 0.020~\text{m}^{-1}$, and (ii) a nonlinear broadening model of the form $\beta_{\mathrm{sens}}^{NL} = 1 + \eta_{br} d + \xi d^2$, where $\xi = 2 \times 10^{-3}~\text{m}^{-2}$ captures higher-order dispersion effects arising from stronger curvature of the MoA profile.
As shown in Fig.~3(c), the qualitative BER ordering with respect to $d_c$ remains preserved across all regimes. In particular, short-to-moderate $d_c$ ($2$–$3$~m) continue to provide the most favorable performance, whereas larger $d_c$ ($5$~m) remain more strongly affected by spreading loss and MoA. When TBE becomes stronger (i.e., larger $\eta_{br}$ or the presence of the quadratic term $\xi d^2$), the BER curves at shorter distances shift toward lower effective SINR thresholds due to the increased temporal separation between the S\&C pulses induced by stronger pulse spreading. However, the degradation at larger distances is not eliminated, confirming that link-budget limitations remain dominant in this regime. Overall, the results indicate that the proposed TBE-aware multiplexing mechanism remains structurally valid beyond the nominal quasi-linear regime and that its performance degrades gradually rather than collapsing under more severe atmospheric broadening conditions. Hence, the JSAC multiplexing principle does not require linear TBE; guard intervals are determined from the duration of received broadened sensing pulse, which can be derived from the HITRAN-based channel response.

\begin{figure*}[ht]
\centering
\subfloat[]{
\includegraphics[width=55mm,height=50mm]{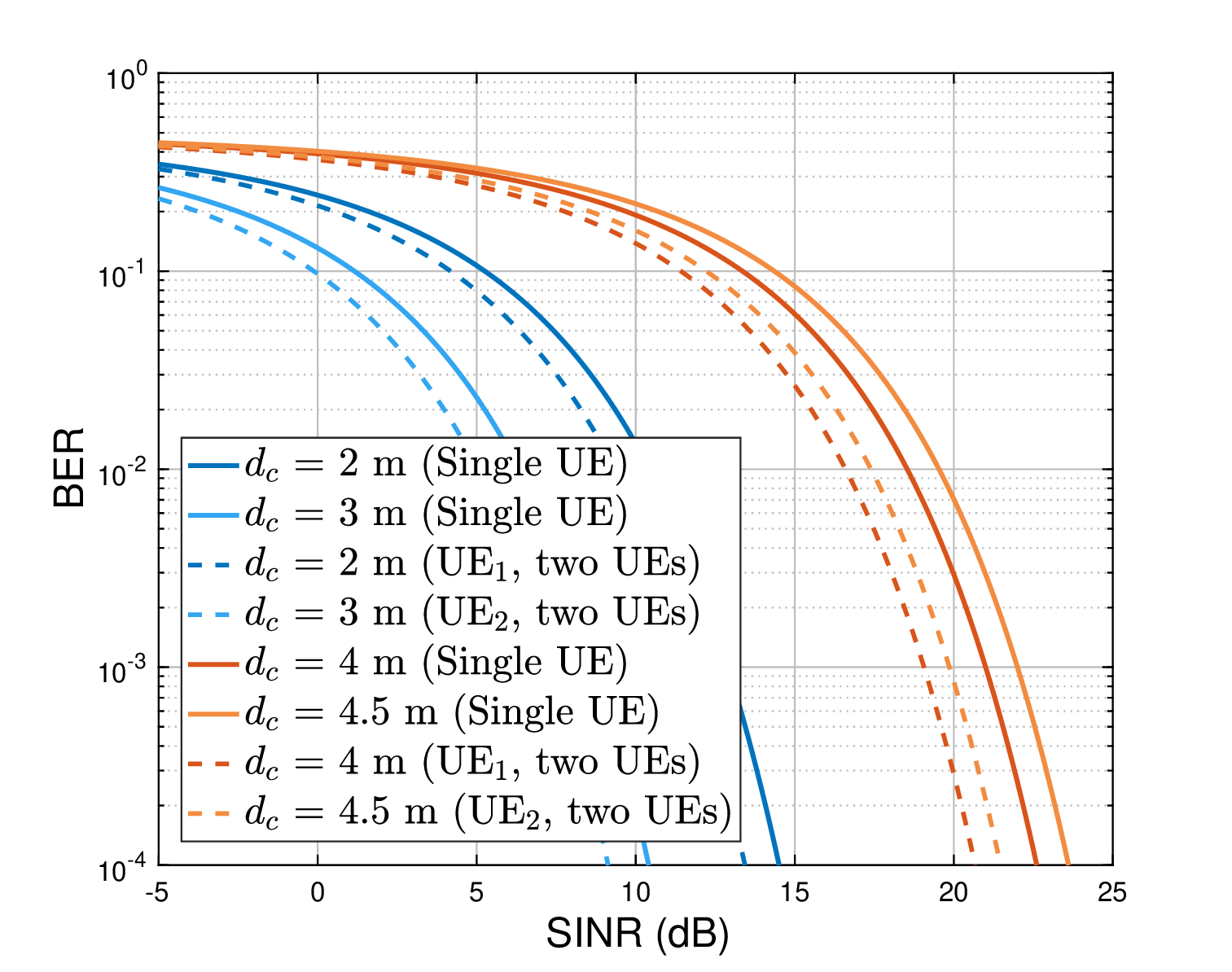}}
\subfloat[]{
\includegraphics[width=55mm,height=50mm]{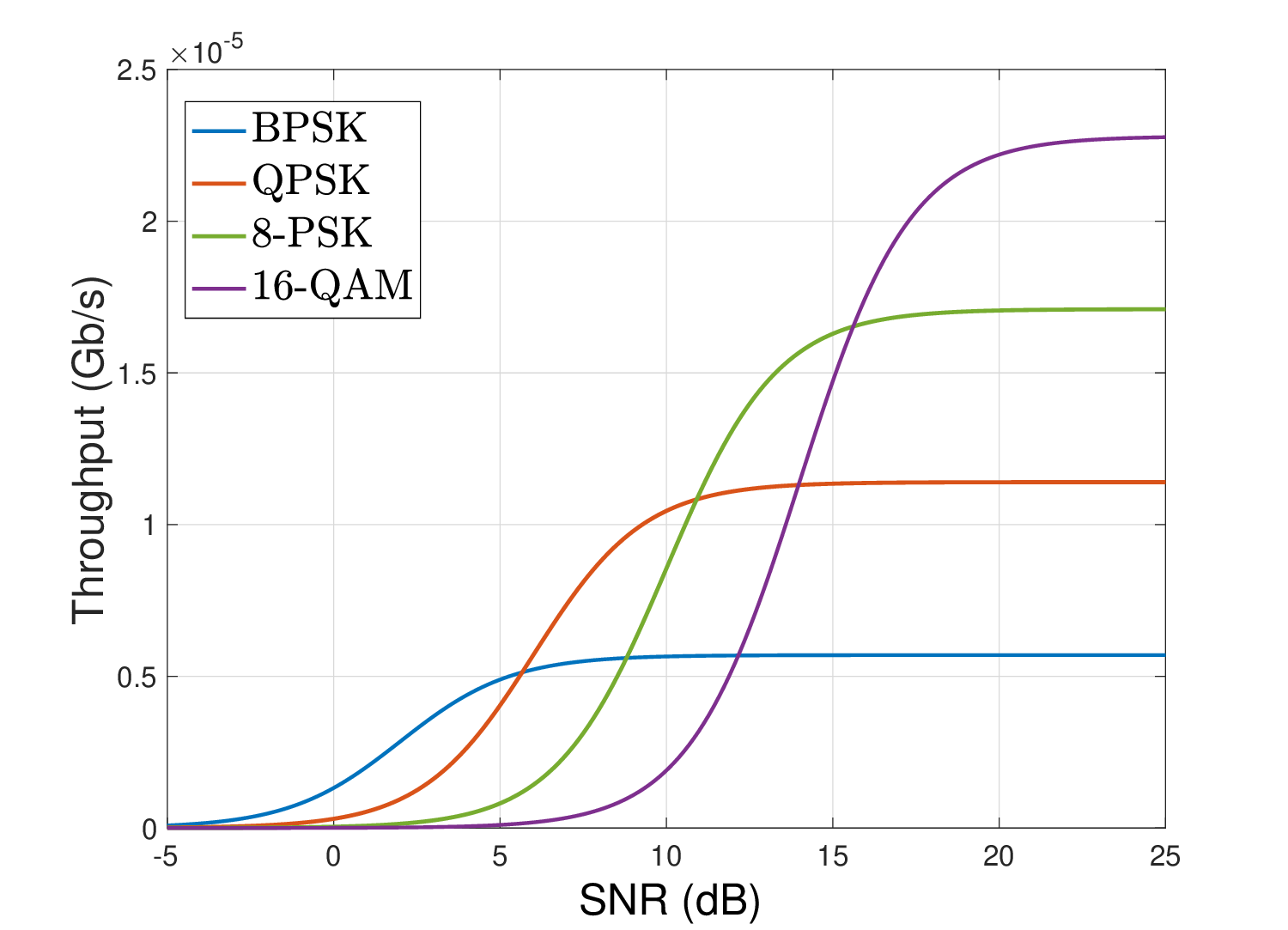}}
\subfloat[]{
\includegraphics[width=55mm,height=50mm]{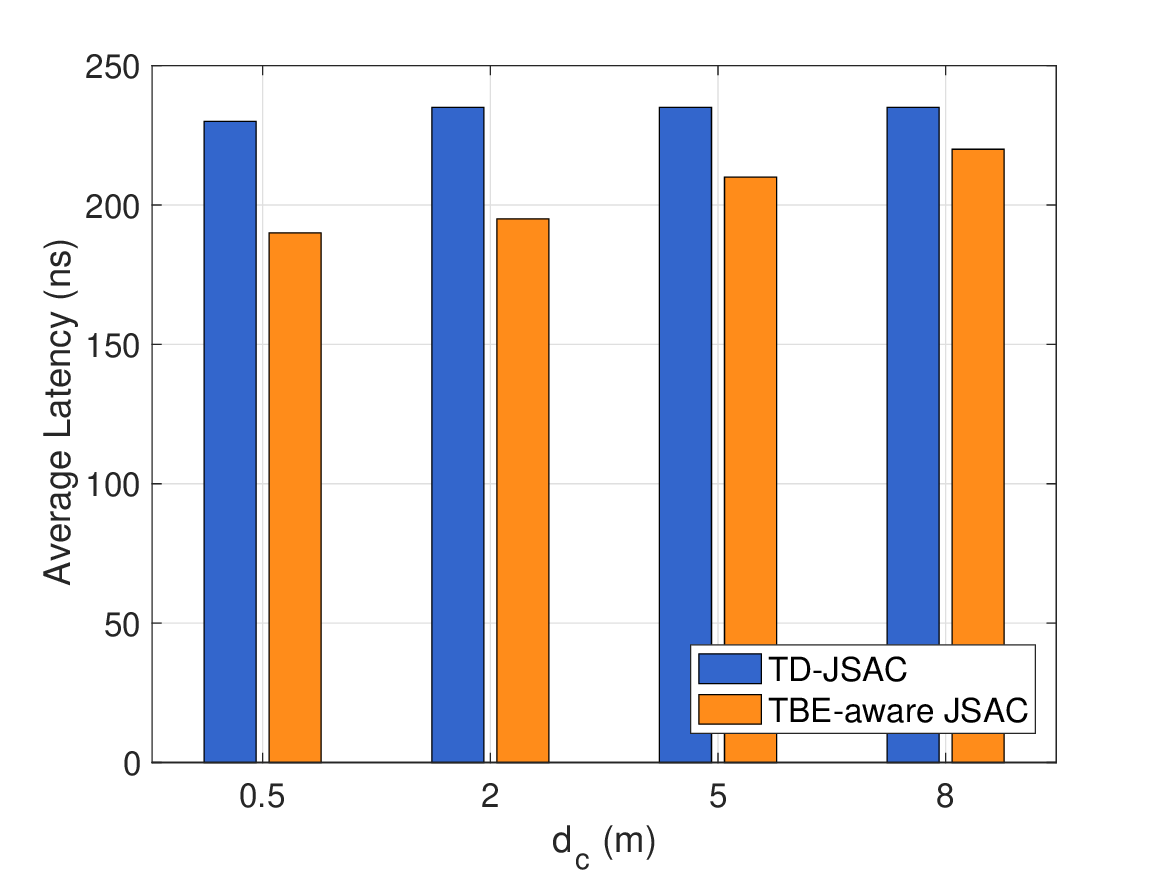}}
\hspace{0mm}
\subfloat[]{ \includegraphics[width=55mm,height=50mm]{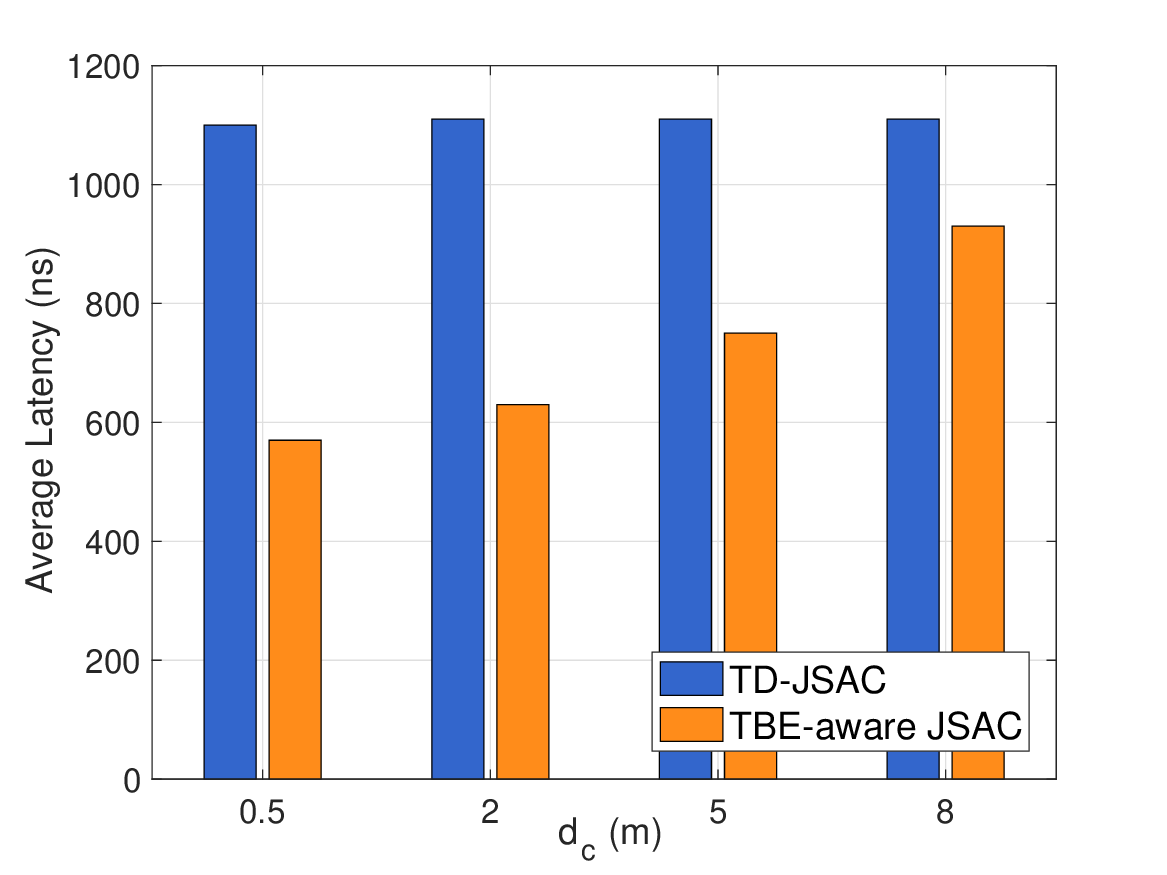}}
\subfloat[]{ \includegraphics[width=55mm,height=50mm]{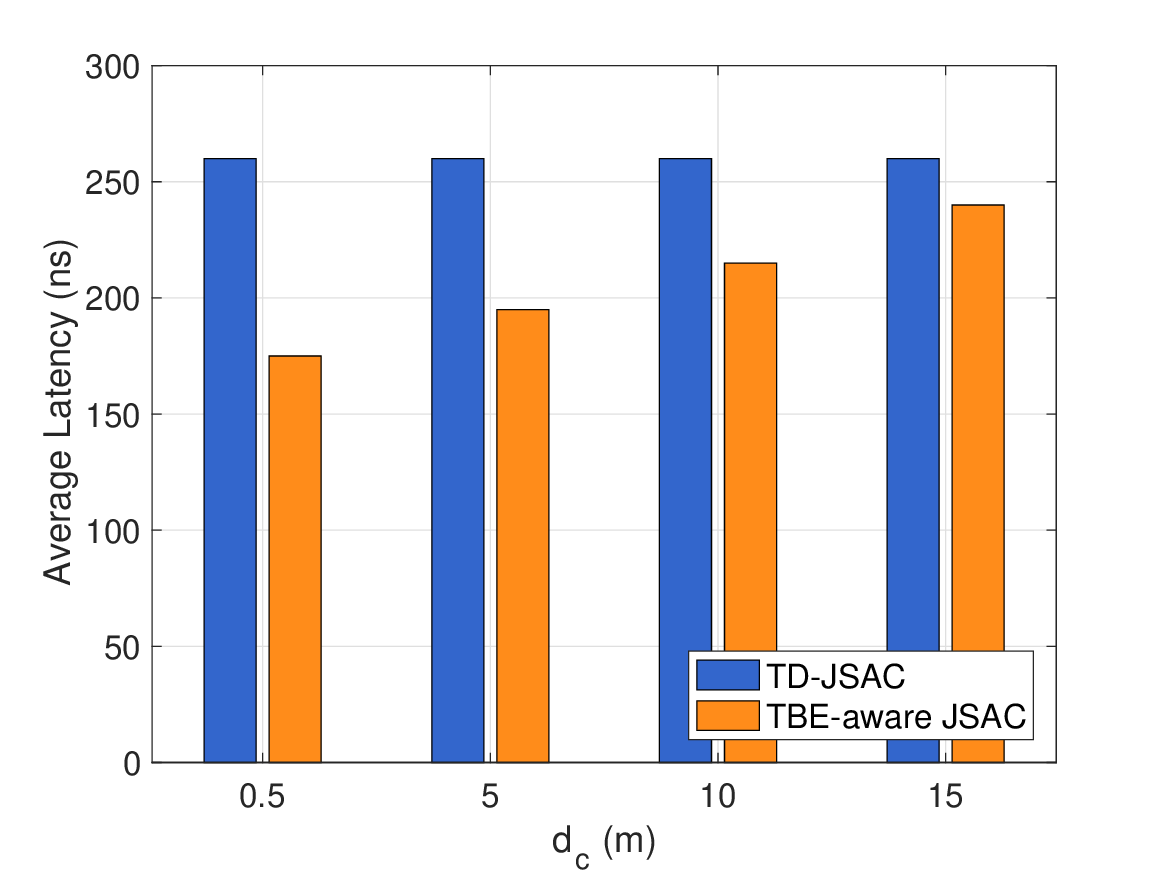}}
\subfloat[]{ \includegraphics[width=55mm,height=50mm]{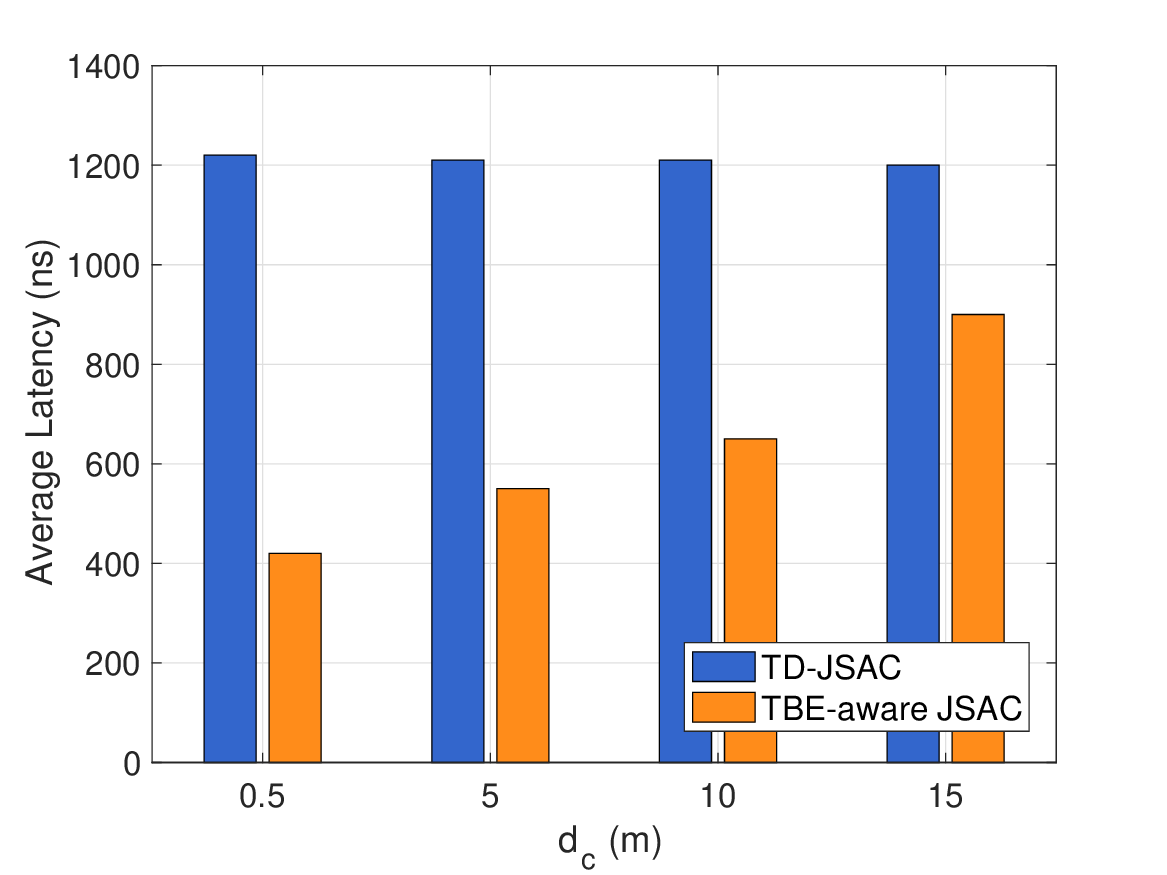}}
\caption{(a) BER versus SINR for two communication users, (b) throughput vs SNR at various modulation orders, (c) $d_c$ vs average latency at $d_s=10$~m and $\rho=0.5$, (d) $d_c$ vs average latency at $d_s=10$~m and $\rho=0.9$, (e) $d_c$ vs average latency at $d_s=20$~m and $\rho=0.5$, and  (f) $d_c$ vs average latency at $d_s=20$~m and $\rho=0.9$.}
\label{simres}
\end{figure*}
\subsection{Probability of Detection}
\par Figure \ref{sens_accuracy}(a) illustrates $P_D$ versus SNR at $\text{Rx}_{\text{sens}}$ for $d_c \in \{3, 4, 5\}$ m. The distance $d_c = 3\,\text{m}$ lies within the optimal distance range (2m--3.5m) yields the best BER performance for communication, as shown in Fig. 4(a), and thus does not require any power adaptation. Notably, the detection performance at far $\text{Rx}_{\text{sens}}$ remains largely unaffected when the $\text{UE}_{\text{comm}}$ moves either closer or farther from it, provided that the $P_c$ remains fixed. This is because communication pulses are significantly attenuated by path loss and TBE, ensuring that their energy at the $\text{Rx}_{\text{sens}}$ remains below the noise floor. However, when power adaptation is applied to compensate for BER degradation at greater distances, the increased $P_c$ raises interference and consequently the noise floor at the $\text{Rx}_{\text{sens}}$, thereby reducing $P_D$, as observed for $d_c = 4\,\text{m}$ and $5\,\text{m}$ with $P_c= 15$ dbm and $20$ dbm, respectively. 
Conversely, when PD-NOMA with SIC is employed, $P_D$ curves follow the same distance and power-dependent trend but show a rightward shift due to residual interference after imperfect SIC. The degradation is minor at $P_c=10$ dbm, moderate at $15$ dbm, and pronounced at $20$ dbm, where the detection threshold moves toward higher SNR. All curves converge at low SNR, confirming the noise-limited regime, while at high SNR, TBE-aware JSAC consistently achieves higher $P_D$ due to its interference-free temporal isolation mechanism, whereas PD-NOMA remains constrained by residual SIC leakage. These results highlight a fundamental S\&C tradeoff, where improving communication robustness through power adaptation can compromise detection accuracy. The proposed design circumvents this tradeoff by avoiding power control altogether, maintaining robust sensing over a range of communication distances without relying on complex interference mitigation at $\text{Rx}_{\text{sens}}$.
\subsection{Sensing Accuracy Analysis}
\par The RMSE is used here as an evaluation metric to quantify sensing estimation accuracy under coexistence conditions, reflecting the impact of noise, TBE, and residual communication interference. The RMSE in estimating the $d_L$ and $v_L$ under interference from a coexisting near $\text{UE}_{\text{comm}}$ are defined as
\begin{equation} \small
\begin{aligned}
\text{RMSE}_{R} &= \sqrt{\frac{1}{K} \sum_{j=1}^{K} \left( \hat{d}_{L_j} - d_{L_j} \right)^2}, \\
\text{RMSE}_{V} &= \sqrt{\frac{1}{K} \sum_{j=1}^{K} \left( \hat{\nu}_{L_j} - \nu_{L_j} \right)^2},
\end{aligned}
\end{equation}
where $K$ is the number of Monte Carlo iterations, and $\hat{d}_{L_j}, \hat{\nu}_{L_j}$ represent the estimated range and velocity for the $j$-th trial, while $d_{L_j}, \nu_{L_j}$ denote their true values.
\par Figure \ref{sens_accuracy}(b) and \ref{sens_accuracy}(c) illustrate the impact of increasing $P_c$ on range and velocity RMSE, respectively, across a range of SNR values at $\text{Rx}_\text{sens}$. Three power levels are considered: $P_c = 10~\text{dbm}$, $P_c = 15~\text{dbm}$, and $P_c = 20~\text{dbm}$, for a fixed $P_s = 25~\text{dbm}$. Increasing $P_c$ introduces interference at $\text{Rx}_\text{sens}$ via the broadened communication pulse, which spectrally overlaps with $z_{\text{sens}}(t)$. This interference causes a notable rise in estimation error, especially in the low-to-medium SNR region. At $P_c = 20~\text{dbm}$, both range and velocity RMSE remain elevated even as SNR increases, indicating that interference from $x_c(t)$ dominates over thermal noise. Conversely, $P_c = 10~\text{dbm}$ ensures minimal interference, resulting in a steep RMSE decline with SNR and the lowest error floor across both metrics. The intermediate case $P_c = 15~\text{dbm}$ shows moderate degradation, highlighting the sensitivity of sensing accuracy to communication power adaptation. 
Incorporating PD-NOMA with SIC (dashed curves) reveals the same qualitative dependence on $P_c$ and SNR, but each PD-NOMA curve lies above and slightly right-shifted relative to its TBE-aware JSAC counterpart due to residual interference following imperfect SIC. The degradation is small for $P_c = 10~\text{dbm}$, moderate at $15~\text{dbm}$, and more pronounced at $20~\text{dbm}$, where the error floor persists even in high-SNR conditions. This behavior confirms that post-cancellation residuals continue to limit sensing precision. Velocity estimation, being inherently more sensitive to spectral leakage and multi-pulse distortion, exhibits greater RMSE separation than range estimation across all power settings. Overall, the results demonstrate that an increase in $P_c$ amplifies interference at $\text{Rx}_{\text{sens}}$, whereas maintaining $P_c = 10$ dBm achieves a balanced operation with robust communication and high-fidelity sensing. As $P_c$ increases, the system transitions from a noise-limited regime ($E_c^{\mathrm{rx}} \ll N_0 B$) to an interference-influenced regime. Consequently, the effective sensing SNR degrades, leading to higher detection thresholds and increased estimation error, as observed in Fig.~4. The TBE-aware JSAC approach outperforms PD-NOMA with SIC by preventing in-slot interference through temporal isolation rather than relying on post-processing interference cancellation.
To assess robustness under increased communication density, a two-user same-frame scheduling scenario is considered, where the communication interval is partitioned as $T_c = T_{c,1} + T_{sep} + T_{c,2}$, with $T_{c,1} = 0.45T_c$, $T_{c,2} = 0.45T_c$, and $T_{sep} = 0.1T_c$. As shown in Fig.~5(a), when users are located within the favorable distance region (e.g., 2 m and 3 m), the additional user introduces only modest BER degradation, indicating that TBE-aware temporal multiplexing can support multiple users with limited performance loss. However, for larger distances (e.g., 4 m and 4.5 m), increased attenuation, molecular absorption, and temporal dispersion shift the BER curves toward higher SINR, with additional degradation due to slot sharing. These results indicate that multi-user operation remains feasible within the favorable propagation region, while performance becomes increasingly constrained outside it. Additionally, user mobility induces gradual variations in TBE that can be accommodated through guard adaptation. At the same time, stricter sensing accuracy requirements primarily affect waveform design and remain compatible with the proposed multiplexing, provided that guard constraints are maintained.
\subsection{Throughput Analysis}
Figure~5(a) illustrates the \textit{throughput--SNR} behavior for the embedded SC modulations (BPSK, QPSK, 8-PSK, and 16-QAM) within the TBE-aware JSAC frame of Fig.~2 using the parameters listed in Table~I. The throughput is calculated using
\begin{equation}
\text{Throughput} ~=~ \frac{\log_2(M)}{T_s + T_g' + T_c},
\end{equation}
where $M$ defines the modulation order (for BPSK $M=2$,  QPSK $M=4$, 8-PSK $M=8$, and 16-QAM $M=16$). The curves increase with SINR as constellation reliability improves and then saturate at the maximum achievable throughput allowed by the system parameters. At low SNR, BPSK provides the most reliable performance ($\approx 0.97$~Gb/s). As the SINR exceeds $\approx 8$~dB, QPSK nearly doubles the achievable throughput ($\approx 1.94$~Gb/s). 8-PSK and 16-QAM further enhance spectral efficiency but require substantially higher SNR for stable demodulation because of tighter symbol spacing and increased error susceptibility. This behavior highlights the classical SNR--complexity trade-off where higher-order modulations improve throughput but become increasingly vulnerable to phase noise, nonlinear distortion, and residual ISI--dominant impairments in THz hardware.
The communication bandwidth in this design remains intentionally narrower than the sensing bandwidth even though both occupy equal temporal durations. The $x_s(t)$ achieves high resolution sensing over a wide 2 GHz band, whereas $x_c(t)$ is a narrowband SC signal optimized for power efficiency, low PAPR, and hardware feasibility. Employing multicarrier schemes such as orthogonal frequency division multiplexing (OFDM) is undesirable in the THz regime due to their high PAPR, strong sensitivity to phase noise and frequency offsets, out-of-band leakage, and complex synchronization and linearization requirements, all of which are amplified at sub-millimeter wavelengths. Consequently, the proposed guard-embedded SC approach achieves a practical balance between energy efficiency, hardware simplicity, and coexistence, allowing throughput enhancement through modulation-order scaling without increasing bandwidth or compromising sensing performance. While modulation order enables throughput scaling, it primarily affects the required SNR threshold for reliable demodulation and does not alter the underlying optimal communication distance. The optimal $d_c$ is governed by the tradeoff between insufficient inner guard at small distances and path loss with molecular absorption at larger distances. Higher-order modulations impose stricter SNR requirements, thereby reducing the feasible operating range, but the location of the optimal region remains unchanged.
\subsection{Latency Discussion}
The average communication latency performance of the proposed TBE-aware JSAC scheme is evaluated and compared against its fixed guard alternative using M/D/1 queuing theory. In the fixed-guard configuration, a constant $T_g$ proportional to the worst-case TBE at the far $\text{Rx}_{\text{sens}}$ results in constant latency regardless of $\text{UE}_{\text{comm}}$ placement, with average delays reaching $230$~ns and $1110$~ns under light ($\rho=0.5$) and heavy ($\rho=0.9$) traffic, respectively, for $d_s = 10$~m. In contrast, the TBE-aware design adapts $T'_g$ according to $d_c$, achieving significant reductions in frame duration and queuing latency. Fig. 5(c) and Fig. 5(d) shows that for $d_s = 10$~m, latency decreases by $19.3\%$ and $49.0\%$ at $d_c = 0.5$~m, and by $16.4\%$ and $43.7\%$ at $d_c = 2$~m, under light and heavy traffic, respectively. These results highlight the consistent benefits even at moderate distances. Additionally, for $d_s = 20$~m, the gains amplify due to the quadratic dependence of queuing delay on frame length: latency reduces by $33.3\%$ and $66.5\%$ at $d_c = 0.5$~m, and remains substantial at $d_c = 5$~m, with $24.8\%$ (light traffic) and $55.6\%$ (heavy traffic) reduction as demonstrated in Fig. 5(e), and Fig. 5(f). Notably, at $d_c = 0.5$~m and $d_s = 20$~m, the absolute latency drop is $811$~ns, enabling over $3300$ additional transmission opportunities per second. The latency gain scales inversely with $d_c$, and the normalized efficiency metric $\zeta$ confirms this spatial advantage. These results validate that the proposed method is especially beneficial for short-range $\text{UE}_{\text{comm}}$s in congested THz scenarios, where exploiting distance-dependent TBE leads to high temporal efficiency without compromising sensing integrity. It is worth mentioning here, that the proposed TBE-aware S\&C frame supports heavy traffic by aggregating symbols and implementing parallel processing. 
\par To clarify the origin of the observed performance gains, the simulation results distinguish two mechanisms. Improvements in communication reliability and sensing accuracy (BER, $P_D$, and RMSE) primarily arise from interference avoidance. In the proposed framework, distance-dependent TBE naturally separates S\&C signals in time, preventing cross-functional interference that would otherwise occur in PD-NOMA. In contrast, the latency reduction reported later in this section results from improved temporal resource utilization. Specifically, adapting the guard interval according to TBE experienced by the communication user shortens the frame duration without altering the fundamental sensing–communication trade-off.
The presented results correspond to a representative short-range THz JSAC configuration with distance asymmetry between S\&C links and moderate atmospheric conditions. While these results validate the feasibility of TBE-aware multiplexing, the conclusions should be interpreted within this regime, and more general scenarios may require adaptive guard design or additional multiplexing strategies.
\section{Conclusion}\label{section5}
This paper introduced a TBE-aware multiplexing framework for JSAC in the THz band. The approach exploits the distance-dependent temporal broadening caused by MoA to reuse the guard interval of a wideband LFM sensing pulse for embedding a low-power single-carrier communication signal. The outer guard is designed for the worst-case broadening at the far sensing receiver, while a short inner guard isolates the nearby communication user from sensing pulse leakage. At the near user, matched filtering enables reliable demodulation without successive interference cancellation, whereas at the far sensing receiver, the broadened and attenuated communication echo remains below the noise floor, ensuring interference-free sensing. Simulation results demonstrate significant performance advantages over PD-NOMA with SIC and a fixed guard counterpart of the proposed scheme. The proposed scheme achieves lower BER, higher probability of detection, and improved range and velocity estimation accuracy, while throughput scales efficiently with modulation order under high SNR. Latency analysis using M/D/1 queuing reveals reductions of up to 66.5\%, confirming enhanced temporal efficiency. The observed gains therefore result from improved interference management and temporal resource utilization rather than from altering the fundamental sensing–communication trade-off. These results establish temporal broadening as an effective physical-domain separator, enabling scalable, low complexity, and interference-free JSAC operation for future THz 6G networks. The proposed framework is particularly suited to THz scenarios exhibiting sufficient distance-dependent temporal asymmetry, typically observed in short-range link. Future directions include dynamic guard adaptation under mobility, multi-target extensions, and evaluation of the proposed framework with alternative waveforms to characterize their interaction with TBE in THz channels and its impact on S\&C coexistence.
\ifCLASSOPTIONcaptionsoff
\newpage
\fi
\bibliographystyle{IEEEtran}
\bibliography{ref}
\end{document}